\documentclass[%
 reprint,
 amsmath,amssymb,
 aps,
]{revtex4-1}

\usepackage{graphicx}

\usepackage{subfigure}
\usepackage{dcolumn}
\usepackage{bm}
\usepackage{hyperref}
\usepackage{amssymb,amsmath,amsthm}
\usepackage[inline]{enumitem}
\usepackage{tasks}

\definecolor{todoGreen}{rgb}{0.0, 0.5, 0.0}
\begin{document}

\title{From canyons to valleys: Numerically continuing sticky hard sphere clusters to the landscapes of smoother potentials}
\author{Anthony Trubiano and Miranda Holmes-Cerfon}
\affiliation{Courant Institute of Mathematical Sciences, New York University, New York, New York 10012, USA}
\date{\today}
\begin{abstract}
We study the energy landscapes of particles with short-range attractive interactions as the range of the interactions increases. Starting with the set of local minima for $6\leq N\leq12$  hard spheres that are ``sticky'', i.e. they interact only when their surfaces are exactly in contact, we use numerical continuation to evolve the local minima (clusters) as the range of the potential increases, using both the Lennard-Jones and Morse families of interaction potentials. 
As the range increases, clusters  merge, until at long ranges only one or two clusters are left. We compare clusters obtained by continuation with different potentials and find that for short and medium ranges, up to about 30\% of particle diameter, the continued clusters are nearly identical, both within and across families of potentials. For longer ranges the clusters vary significantly, with more variation between families of potentials than within a family. We analyze the mechanisms behind the merge events, and find that most rearrangements occur when a pair of non-bonded particles comes within the range of the potential.  An exception occurs for nonharmonic clusters, those that have a zero eigenvalue in their Hessian, which undergo a more global rearrangement. 

\end{abstract}

\maketitle


\section{Introduction}

Metastable states of a system of interacting particles determine much of the system's behaviour, yet they can be difficult to study because they are sensitive to the interaction potential between the particles \cite{potentialsAS,potentials2,Wales:2003}. 
For mesoscale particles like colloids, the interaction potential is not always well known, because it depends on a combination of factors 
 that occur on a much smaller scale than the particles, 
such as electrostatic interactions, van der Waals interactions, the presence of impurities in solution, complex surface interactions created by tethered polymers, and other physical effects. Experimentally, the interaction potential is hard to measure because the particles typically interact over a distance much smaller than their diameters \cite{manoColloid}. To model attractions between such particles one typically chooses an interaction potential from a canonical family such as Morse, Lennard-Jones, or square-well potentials, and chooses parameters to fit aspects of experimental data. Yet, even for these families of potentials it is not known how sensitive the metastable states are to the choice of potential or parameters, nor how the metastable states for different choices are related to each other.

Conveniently, it has been shown that when particles have short or even medium-ranged attractive interactions,  aspects of their phase behaviour are insensitive to the exact shape of the interaction potential, but rather depend on a single parameter characterizing the potential, the second virial coefficient \cite{noro}. The same is true of the set of metastable states, provided the range is short enough  \cite{FELAHS,Perry:2015ku,singularC}. 
This observation has motivated studying the energy landscape in the \emph{sticky limit} when the range of the potential goes to zero and the depth goes to infinity, in such a way that the partition function approaches a delta-function at the point of contact \cite{BaxterSHS,HCgeometrical}. In this limit, the metastable states of a system of $N$ identical spherical particles are the set of sphere packings that have the most pairs of spheres in contact. Finding these Sticky Hard Sphere (SHS) clusters is a problem in geometry that has been addressed using several  techniques, both analytical and numerical \cite{ArkusSP,Hoy:2012cr,Hoy:2015hz,clusters}, and the resulting data has given insight into a variety of physical properties of mesoscale particles \cite{PatrickRoyall:2008fz,Malins:2009dt,FELAHS,Perry:2015ku}. 
However, real experimental colloidal systems do not always lie close enough to the sticky limit for it to be quantitatively accurate, and discrepancies from the predictions of the sticky limit have been observed even for systems as small as $N{=}8$ particles \cite{FELAHS}. 

We seek to understand how sensitive the metastable states are to the choice of potential when a system is near, but not exactly at, the sticky limit. Starting with the sticky-sphere landscape, which is thought to be the most rugged and to contain the most local minima, we apply numerical continuation to follow local minima as we slowly increase the range within a family of potentials for systems of $6\leq N\leq12$ spheres.
This procedure finds most of the local minima for smooth potentials, and in particular all the known deep local minima. 
We compare clusters that come from the same SHS cluster using different potentials, and find that clusters are nearly for short ranges, up to about $30\%$ of particle diameter, but vary significantly for longer ranges. 
We keep track of bifurcation events, where local minima split, merge, or disappear, 
and show that most bifurcations involving rearrangements occur when a pair of non-bonded particles comes within the range of the potential. An exception are bifurcations involving \emph{nonharmonic} SHS clusters (those whose Hessian has a zero eigenvalue which does not extend into a finite floppy mode), which undergo a more global rearrangement whose location cannot be predicted from the starting SHS cluster. 

Our study builds on others that have examined how energy landscapes vary as the range of the pair potential is varied. Wales \cite{walesFold} argued that catastrophe theory gives a quantitative relationship between local minima and the nearest saddle points when  close to a bifurcation, and empirically showed this relationship holds reasonably well even away from the bifurcation. 
Trombach et al \cite{SHStoLJ} performed a local optimization in a Lennard-Jones$(m,n)$ potential with varying $m,n$ (varying range) at fixed energy, using SHS clusters as an initial condition for the optimization, and found most of the local minima on the Lennard-Jones landscapes; they showed the ones not found were from a small set of initial ``seeds''. 
Trombach et al \cite{Trombach:2018ie} followed a similar approach to study the ``kissing problem,'' which asks how to arrange 12 spheres on the surface of a central sphere, in a family of Lennard-Jones potentials. 
The latter two approaches are the closest to ours; however these studies performed a one-step optimization for each value of range, hence could only compare the number of clusters found. In contrast, we vary the range parameter slowly, using the previously-found cluster as the next initial condition, so we can additionally find and study bifurcations.

\section{Methods}

\begin{figure}[t]
\includegraphics[width=0.42\textwidth]{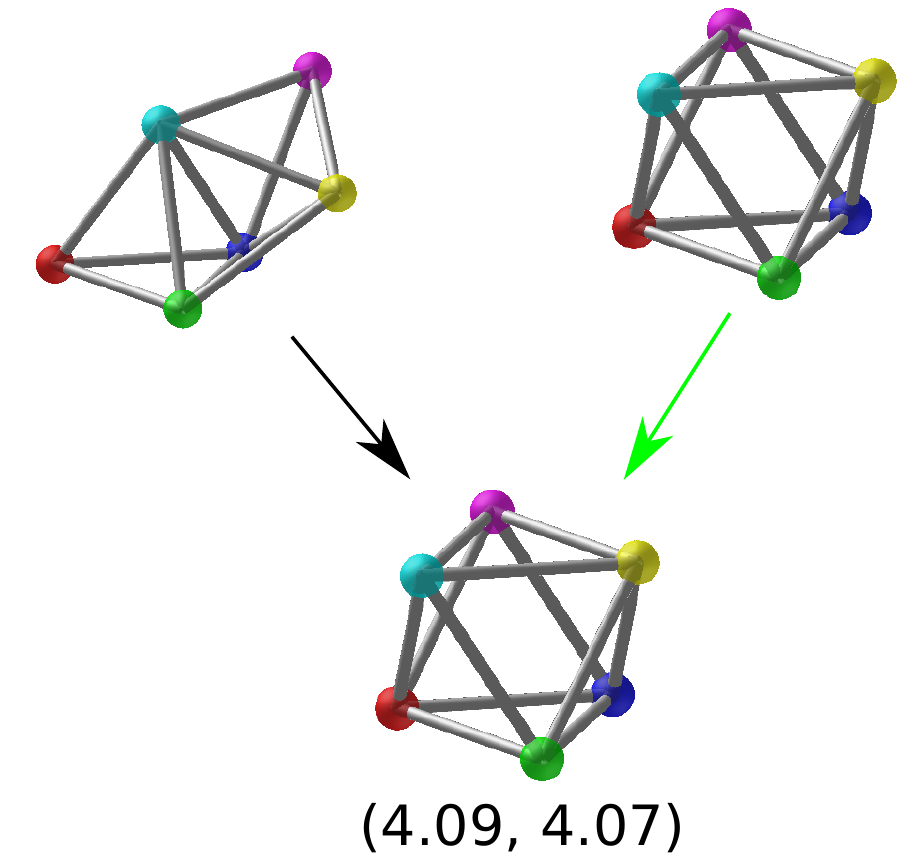} 
\centering
\caption{Merging tree for $N=6$ particles.  The top row of the tree contains all SHS clusters (left = polytetrahedron, right = octahedron), and the tree follows unique clusters through the continuation process. The ordered pair $(\rho,m)$ beneath the final cluster give the range values for which the clusters merged for the Morse and Lennard-Jones potentials, respectively. The polytetrahedron suddenly rearranges and becomes identical to the octahedron. The octahedron remains the same throughout the continuation procedure. The topology of the trees is the same for each choice of $\kappa$ and for each potential. Green arrows denote a smooth transition between clusters; black arrows indicate a cluster rearranged non-smoothly. Bars indicate an inter-particle distance less than or equal to $1$. 
} 
\label{N6Merge}
\end{figure}

We begin with a set of SHS clusters that is thought to be nearly complete, likely  missing only a small number of high-energy, nonharmonic clusters \cite{clusters}. This set was produced using an algorithm to enumerate clusters that starts with a single rigid cluster, breaks a contact, and follows the one-dimensional transition path until a new contact is formed, producing a new rigid cluster. Iterating over all bonds and then all rigid clusters in the evolving list gives the dataset \cite{clusters}. 
We consider how each of these clusters evolves as we slowly smooth out the pair potential into either a Morse or Lennard-Jones potential, given respectively by
\begin{align} \label{Potentials}
U_{M}(r)&=E\left(e^{-2\rho(r-d)}-2e^{-\rho(r-d)}\right),\\
U_{\text{LJ}}(r)&=\frac{E}{m}\left(m\left(\frac{d}{r}\right)^{2m}-2m\left(\frac{d}{r}\right)^m\right). 
\end{align}
Here $r$ is the inter-particle distance, $E>0$ is the well depth, $\rho$ and $m$ are parameters governing the inverse range of the potential, and $d$ is the equilibrium bond distance; we choose units so that $d=1$.

We perform continuation on the set of SHS clusters as follows. We set the initial range parameters to be $\rho=m=50$, and choose a corresponding energy parameter $E$. At each step of the continuation we decrease the range parameter by $0.01$, which slowly increases the range, and we update $E$ in a way we describe momentarily. We then minimize the potential energy (either $U_{M}$ or $U_{\text{LJ}}$) under the new parameter values using the conjugate gradient algorithm, with the clusters obtained at the previous step as an initial condition (see Appendix \ref{CG} for details.) These steps are repeated until the range parameter becomes $1$. 

Figure \ref{N6Merge} illustrates the continuation procedure using a Morse potential for $N=6$. There are two SHS clusters, which we will call the polytetrahedron and the octahedron. The octahedron remains the same throughout the continuation process. The polytetrahedron remains nearly the same for range parameters $\approx\rho>5$, with inter-particle distances varying by less than about $3\%$. As $\rho$ decreases below 5, the central bond in the polytetrahedron (between the blue and cyan particles) stretches slightly, and then the outer bonds (pink-yellow and red-green) stretch slightly too, by about 10\%, as the red and pink particles come slightly closer together. At $\rho=4.09$, the cluster suddenly rearranges, in one optimization step, when the red and pink particles come together to form the octahedron. For $\rho < 4.09$, there is only one Morse cluster, which is identical to the original octahedron.


During the optimization step, it is possible to reach a saddle point rather than a local minimum. This possibility is checked by computing the eigenvalues of a Hessian matrix. If a negative eigenvalue is found, a re-optimization procedure is performed in which the critical point is displaced in both directions along the corresponding eigenvector to obtain new starting points for the conjugate gradient algorithm. The algorithm could then produce two distinct local minima and we keep track of any such splitting events.  

After constructing these lists of clusters, we compare each cluster pairwise to determine whether they are unique up to translations, rotations, and permutations (see Appendix \ref{testSame} for details.) If two clusters are not unique we say their ``parent'' clusters from the previous step have ``merged.''   
For each family of potentials and each choice of energy parameters, we construct a bifurcation diagram showing how clusters merge and split as a function of the range parameter. It is this bifurcation diagram that we study in the remainder of the text.

During the continuation we must decide how to vary the energy $E$ as a function of the range parameters $\rho,m$. We don't believe this choice is terribly important to the results. We decided to try to keep the equilibrium distributions for systems along a single continuation path roughly comparable, so we chose $E$ so the partition function for a bond between a pair of particles on a line remains approximately constant.
That is, define the one-dimensional pair partition function to be $\kappa = \int_0^{r_c} e^{-\beta U_{(\cdot)}(r)}dr$, where $r_c$ is a cutoff beyond which $U_{(\cdot)}\approx 0$ and $\beta = (k_BT)^{-1}$ is the inverse of Boltzmann's constant times the temperature. 
The partition function $\kappa$ is proportional to the time that a pair of one-dimensional particles in isolation spend within a distance of $r_c$ of each other, when the system is in thermal equilibrium. 
The parameter $\kappa$ is known as the 
 \emph{sticky parameter}, because it measures how sticky the particles are -- larger $\kappa$ means the particles spend more of their time nearly in contact with each other \cite{HCgeometrical}. It can be shown (see Appendix \ref{stickyDeriv}) that the sticky parameter is a linear function of the second virial coefficient, $B_2$, which  characterizes thermodynamic properties of a system through the Law of Corresponding States \cite{noro}. 
 
The sticky parameter can be approximated using Laplace asymptotics for a potential with a deep and narrow attractive well 
as $\kappa=\sqrt{2\pi}e^{-\beta U_{(\cdot)}(d)} / \sqrt{\beta U''_{(\cdot)}(d)}$ (see Appendix \ref{stickyDeriv} for details).
Evaluating this expression for the Morse and Lennard-Jones potentials and non-dimensionalizing the energy using units of $\beta$ (so we may set $\beta=1$ in the above formulas) gives
\begin{equation} \label{StickyM}
\kappa_M(\rho,E)= \sqrt{\frac{\pi}{E\rho^2}}e^{ E}, \quad \kappa_{\text{LJ}}(m,E)=\sqrt{\frac{\pi}{ Em^2}}e^{ E}.  
\end{equation}
Although these expressions are meaningful only for very short-ranged potentials, we use them to determine the relationship between the range parameters $\rho,m$ and the energy parameter $E$ at all ranges. 
Notice that the parameters $\rho, m$ both measure the inverse range, and they appear in the formulas above in the same way, so we will use these parameters interchangeably hereafter. 

We perform continuation for each of $U_{M}$, $U_{\text{LJ}}$, and for each of three different values of the sticky parameter, $\kappa_{\text{LOW}}=23.4$, 
$\kappa_{\text{MED}}=49.5$, and $\kappa_{\text{HIGH}}=100.4$. At each step of the continuation we solve \eqref{StickyM} for $E$ using Newton's method. This gently increases $E$ as $\rho$ or $m$  decreases (the range increases.)

\section{Results}

\subsection{Completeness of the Set of Continued Clusters}

\begin{figure}[t]
	\includegraphics[width=0.42\textwidth]{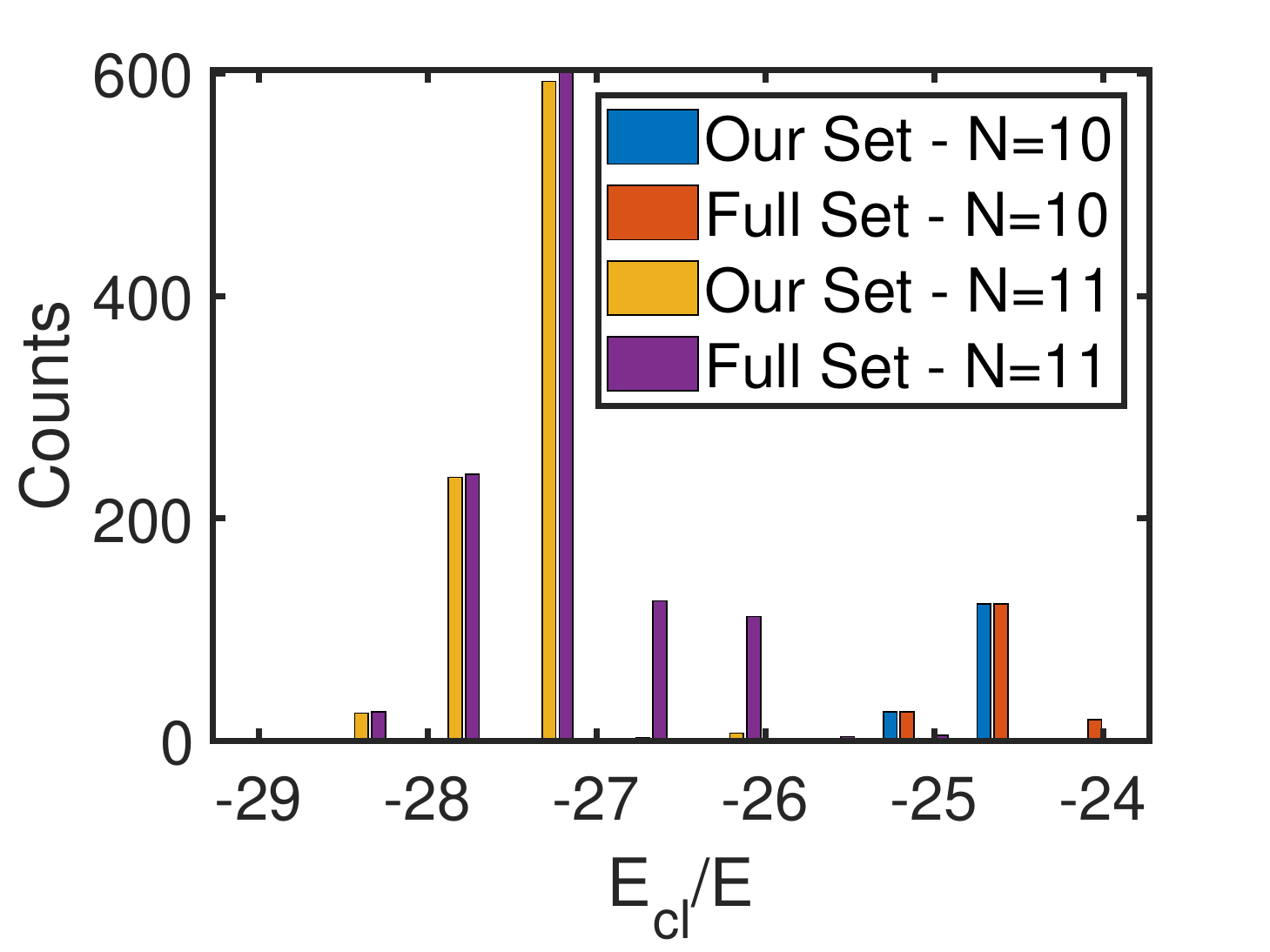} 
	\centering
	\caption{Histogram of the cluster energies, $E_{cl}$, scaled by the well-depth, $E$, for $N=10$ and $N=11$. We compare the energies of the clusters found at $\rho=30$ for the Morse potential with $\kappa = \kappa_{MED}$ using continuation, against the set of all known clusters.  Most of the clusters missed using continuation are high energy. 
	}
	\label{energy}
\end{figure}

\begin{table}[ht] 
\begin{tabular}{l c | c c c | c c c}
 $N$ & $C_N$ & $\left| \text{SHS}\rightarrow M_{30}\right|$ & $\left|M_{30}\right|$ & $\Delta_{30}$ & $\left| \text{SHS}\rightarrow M_{6}\right|$ & $\left|M_{6}\right|$ & $\Delta_6$\\
\hline
6	 & 2 & 2 & 2 & 0 & 2 & 2 & 0\\
7    & 5 & 4 & 4 & 0 & 4 & 4 & 0\\
8    & 13 & 10 & 10 & 0 & 8 & 8 & 0\\
9    & 52 & 30 & 31 & 1 & 17 & 19 & 2\\
10   & 263 & 151 & 170 & 19 & 57 & 61 & 4\\
11   & 1659 & 866 & 1127 & 259 & 161 & 170 & 9 \\
$12^*$ & 11980 & 5684 & 8059 & 2375 & 489 & 506 & 17\\
\end{tabular}
\caption{Number of unique SHS clusters with $N$ particles, $C_N$,  as well as the number of Morse clusters found through the continuation procedure, $\left| \text{SHS}\rightarrow M_\rho\right|$, and the total number of Morse clusters, $\left|M_\rho\right|$, for range parameters $\rho=6,30$. The difference between the continued and complete sets, $\Delta$, is also reported. The continued clusters were generated using sticky parameter $\kappa_{\text{MED}}$. The $*$ indicates that a heuristic algorithm was used to determine uniqueness of clusters, described in the Appendix. }
\label{completeness}
\end{table}

First we ask whether this continuation procedure produces all the local minima for a given landscape.   
We compare the set of Morse clusters obtained by continuation for $\rho=30$ and $\rho=6$ to the local minima found by a basin-hopping technique  in \cite{morseClusters}. The number of unique local minima in each set is given in Table \ref{completeness}. Our method finds all local minima in the basin-hopping dataset for $N\leq 8$, but for larger $N$ it misses a few. Upon inspection, the unmatched clusters are mostly high energy clusters: each unmatched cluster has energy greater than $-(3N-6)E$ and usually close to $-(3N-7)E$, whereas a typical matched cluster has energy between $-(3N-5)E$ and $-(3N-6)E$. 
Figure \ref{energy} compares the energy distributions of the clusters we find at $\rho=30$ and the basin-hopping data. 
A smaller fraction of clusters are missing at longer range: at $\rho=30$ the method missed 11\%, 23\% for $N=10,11$ respectively, whereas for $\rho=6$ the method missed 6.5\%, 5.3\% respectively. 
The continuation procedure did not find any structures that were not present in the basin-hopping data set. 

Our results are comparable to those of Trombach et al. \cite{SHStoLJ}, which computed  
Lennard-Jones clusters with $m{=}6,E{=}1$ ($\kappa{=}0.8$) by performing a one step optimization from a SHS cluster. For $N{=}10,11$ their method failed to find 2/64 (3.1\%) and 5/170 (2.9\%) for $N{=}10,11$ respectively, slightly smaller numbers than ours. 
They also found that most missing clusters were high energy.

If missing clusters are high energy, this suggests that as the range increases, local minima are created on the flat, higher energy parts of the sticky-sphere landscape, from configurations corresponding to floppy clusters with one or more internal degrees of freedom. Such creation of local minima cannot be detected by our procedure. 
Because we obtain better agreement at longer ranges, we hypothesize that these high-energy local minima disappear at larger ranges.  
Typically one is interested in low-energy minima, so we feel confident using our dataset going forward to understand bifurcations in the low-energy parts of the landscape. 


\begin{figure}[t]
\centering  
\subfigure[]{\includegraphics[trim={0.1cm 0cm 0.1cm 0cm},clip,width=0.48\linewidth]{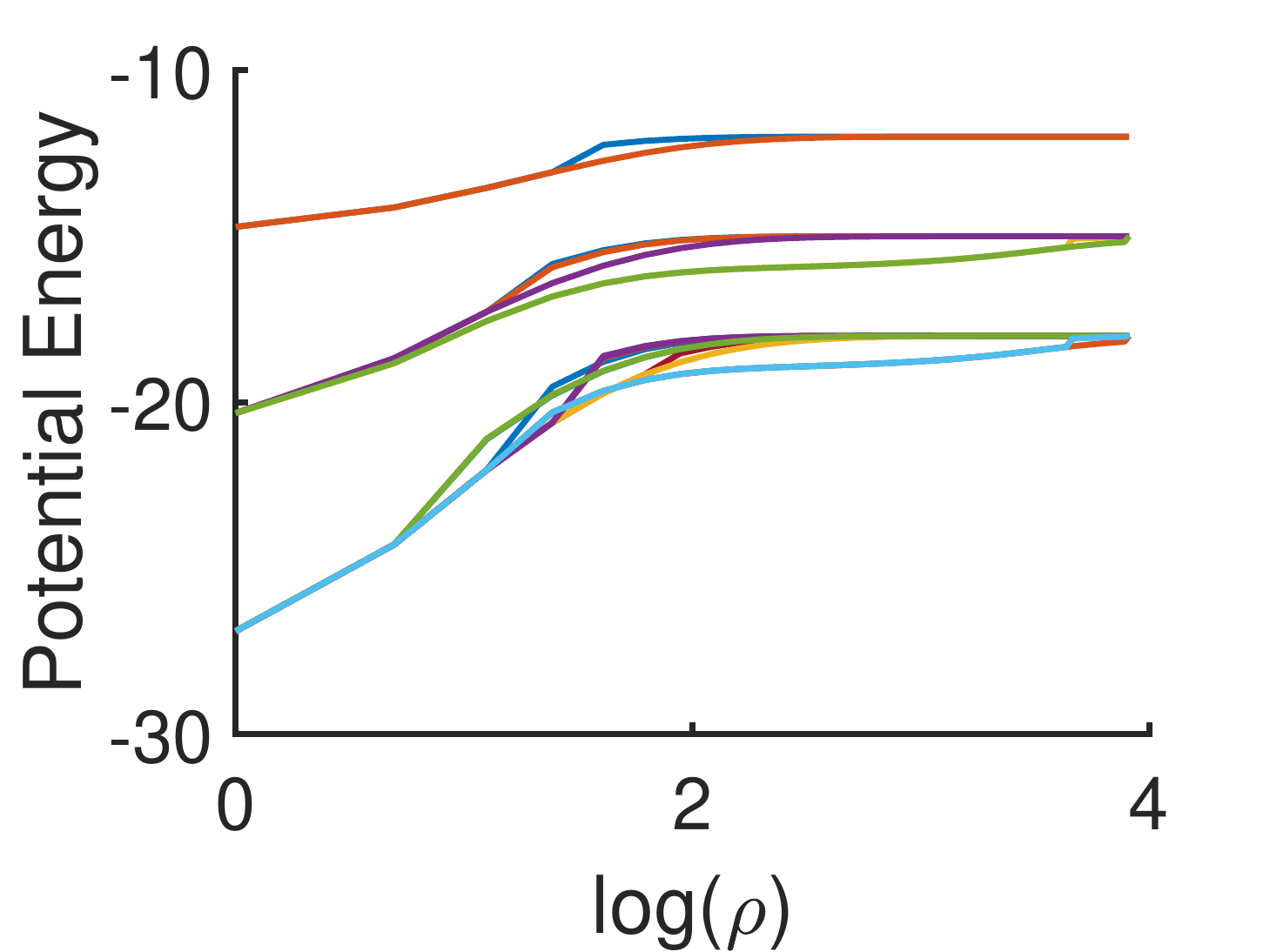}}
\subfigure[]{\includegraphics[trim={0.1cm 0cm 0.1cm 0cm},clip,width=0.48\linewidth]{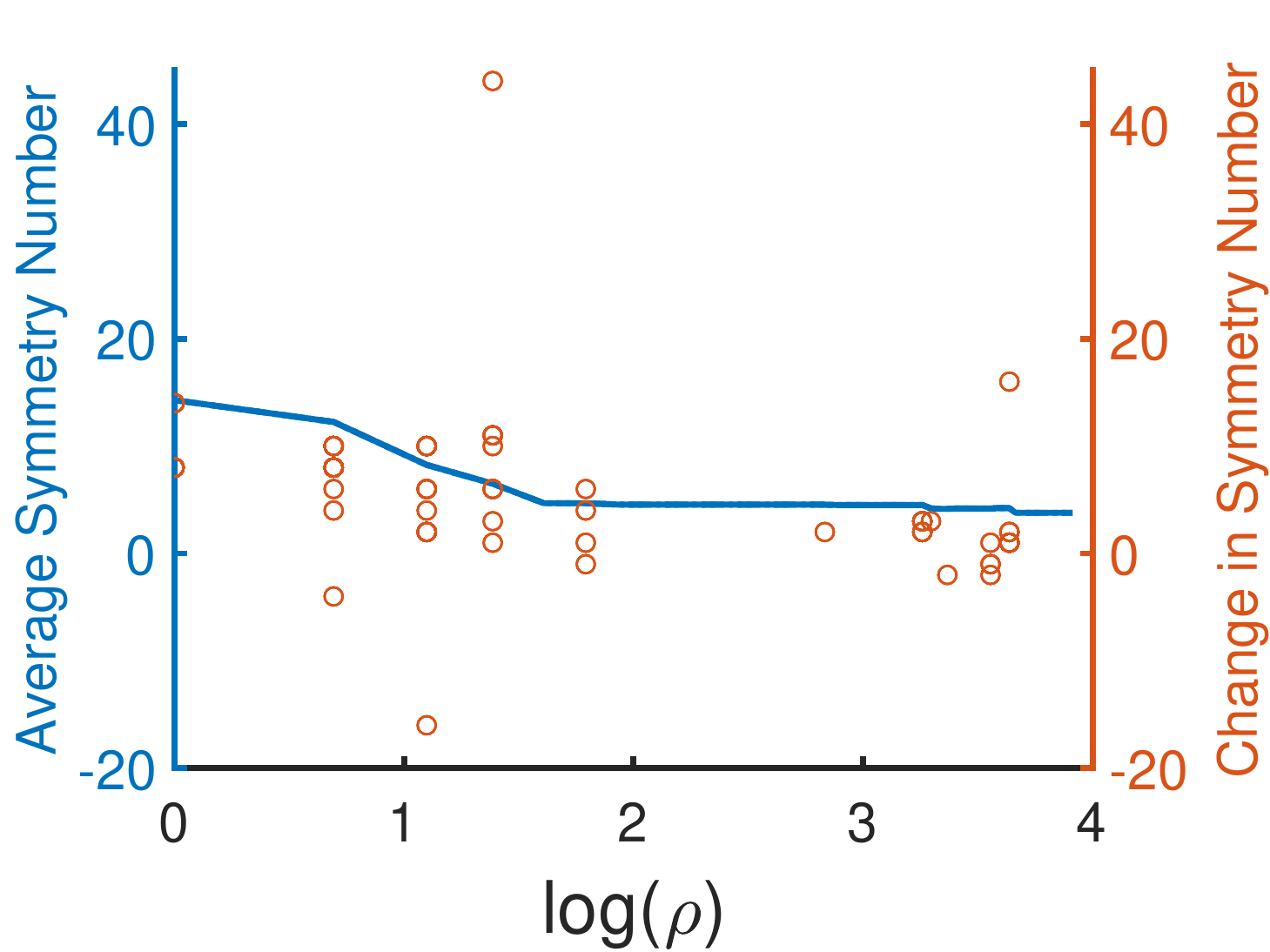}}
\caption{(a) Average potential energy of all clusters as a function of $\log(\rho)$ for $N=(6,7,8)$ from top to bottom, respectively, 
and (b) the average symmetry number of all $N\leq 9$ clusters as a function of $\log(\rho)$, as well as a scatter plot of the change in symmetry number for each individual cluster as a function of $\log(\rho)$. Negative values for the scatter plot mean a cluster's symmetry decreased when it merged. The clusters used for these calculations were generated using the Morse potential and $\kappa = \kappa_{MED}$. 
} 
\label{statGeo}
\end{figure}

As a brief application of our nearly-complete data we show how statistical and geometric properties of the clusters evolve as a function of the range. Figure \ref{statGeo} shows how the Morse potential energy and symmetry number (order of the point group) vary with range. 
The average potential energy decreases as the range of the potential increases, presumably because particles can interact attractively with neighbours that are farther away. The average symmetry number increases as the range increases: clusters for longer-ranged potentials are more symmetric, on average. Interestingly, the symmetry number does not always increase monotonically following a particular cluster; occasionally a cluster acquires lower symmetry as the range increases. 


\subsection{Visualizing bifurcations in the energy landscape}

\begin{figure}[t]
\includegraphics[width=0.48\textwidth]{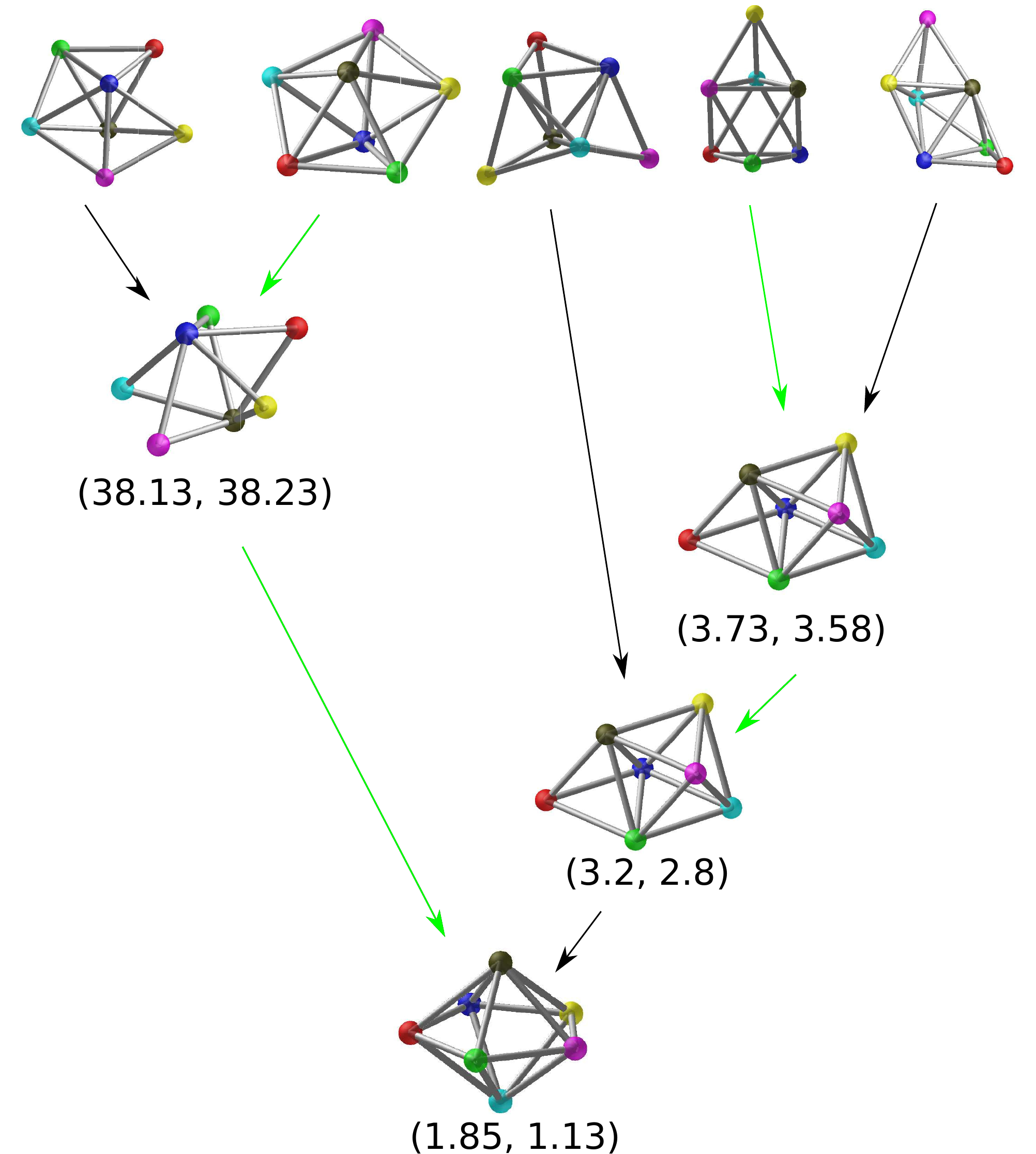}
\centering
\caption{Merging tree for $N=7$, with Morse clusters plotted at nodes of the tree. The top row of the tree contains all SHS clusters and the tree follows unique Morse structures through the continuation process. The topology of the trees is the same for each choice of $\kappa$ and for each potential. The ordered pairs $(\rho,m)$ beneath each cluster give the range value for which the merge happened for Morse and Lennard-Jones potentials respectively. Green arrows denote a smooth transition between clusters; black arrows indicate a cluster rearranged non-smoothly. Bars indicate an inter-particle distance less than or equal to $1$. 
}
\label{N7Merge}
\end{figure}

Next we examine bifurcations in the energy landscape, and compare bifurcation diagrams for different potentials and parameters. 
There are two kinds of bifurcations we can detect: merging events, when two or more local minima become the same cluster (like when the poytetrahedron and octahedron merge in Figure \ref{N6Merge}), and splitting events, when one local minimum splits into two or more.

We find many merging events as the range parameter decreases. 
Interestingly, we find no nonisomorphic splitting events.
Splitting events are possible when a cluster hits a saddle point in the optimization. This only happened when we tracked nonharmonic clusters, the smallest of which occurs at $N=9$. These clusters hit a saddle point initially, and then continued to hit saddle points every so often until $\approx \rho < 30$. However, every time we hit a saddle point and searched both directions of the negative eigenvector, we always found two local minima that were the same up to a rigid rotation or a permutation of the particle labels, hence, which are identified as the same cluster. 
This result was unexpected -- our original hypothesis was that nonharmonic clusters would lead to nontrivial splitting events -- and we do not have an explanation for why we see none.
 
Because we find only merging events, our data can be represented as a graph with a tree structure. The top row contains all SHS clusters, and clusters which merge are connected at a branch in the tree, with a node representing the cluster they merge into. When two or more clusters merge into one, we call the clusters before the merge the ``parents'' and the cluster just after the merge the ``child.''

The simplest merging tree is the one for $N=6$ (Figure \ref{N6Merge}.)
The topology of the merging tree is the same for all potentials and all parameter values we considered. The single merge event occurs at slightly shorter range  for the Morse potential ($\rho=4.09$) than for the Lennard-Jones potential ($\rho=4.07$). 


For $N=7$ the merging trees continue to have the same topology for all potentials and parameters, while the range at which the merges occur depends on the potential  (Figure \ref{N7Merge}.) 
Lennard-Jones clusters usually merge at smaller values of the range parameter (longer range) than Morse clusters, although for the single merge at large range parameter ($\rho=38.13,m=38.23$), the Lennard-Jones cluster merged first. 
Similar observations hold for $N=8$ (see Appendix \ref{n8trees} Figure \ref{N8Merge}): the trees are all topologically the same, but the ranges where merges occur differ slightly between the two potentials.


How do the clusters of $N=7$ merge? In each merge there is always one parent cluster that does not change, and one that rearranges significantly. 
We speculate on the mechanism of the more dramatic rearrangement by inspecting the clusters. 
In the first merge in Figure \ref{N7Merge}, the red and yellow particles of cluster 1 (counting clusters from the left) begin to interact with each other and pull toward the center of mass, pushing the central particles apart to form the more symmetric child cluster. This child cluster has a ring of outer particles that are slightly farther apart than they were in the smoothly-varying parent SHS cluster, cluster 2, presumably because spreading apart the ring allows the central blue and brown particles to come closer together. The second and third merges are like the $N=6$ merge: clusters 3 and 5 each contain a polytetrahedron, which suddenly rearranges into an octahedron, a sub-structure of cluster 4 that they each merge with. 
The merges occur at slightly different ranges; one reason could be that clusters 4 and 5 in the second merge both have six symmetries, so they are more similar to begin with than clusters 3 and 4  in the third merge, which have two and six symmetries respectively. 
In the final merge, the square base of the rightmost cluster absorbs the red particle to become a pentagon. Overall, only one SHS cluster, cluster 2, evolves smoothly throughout the whole continuation process; this cluster happens to be the one with the most symmetries initially.

\begin{figure}[t]
\centering
\includegraphics[height=0.17\textheight]{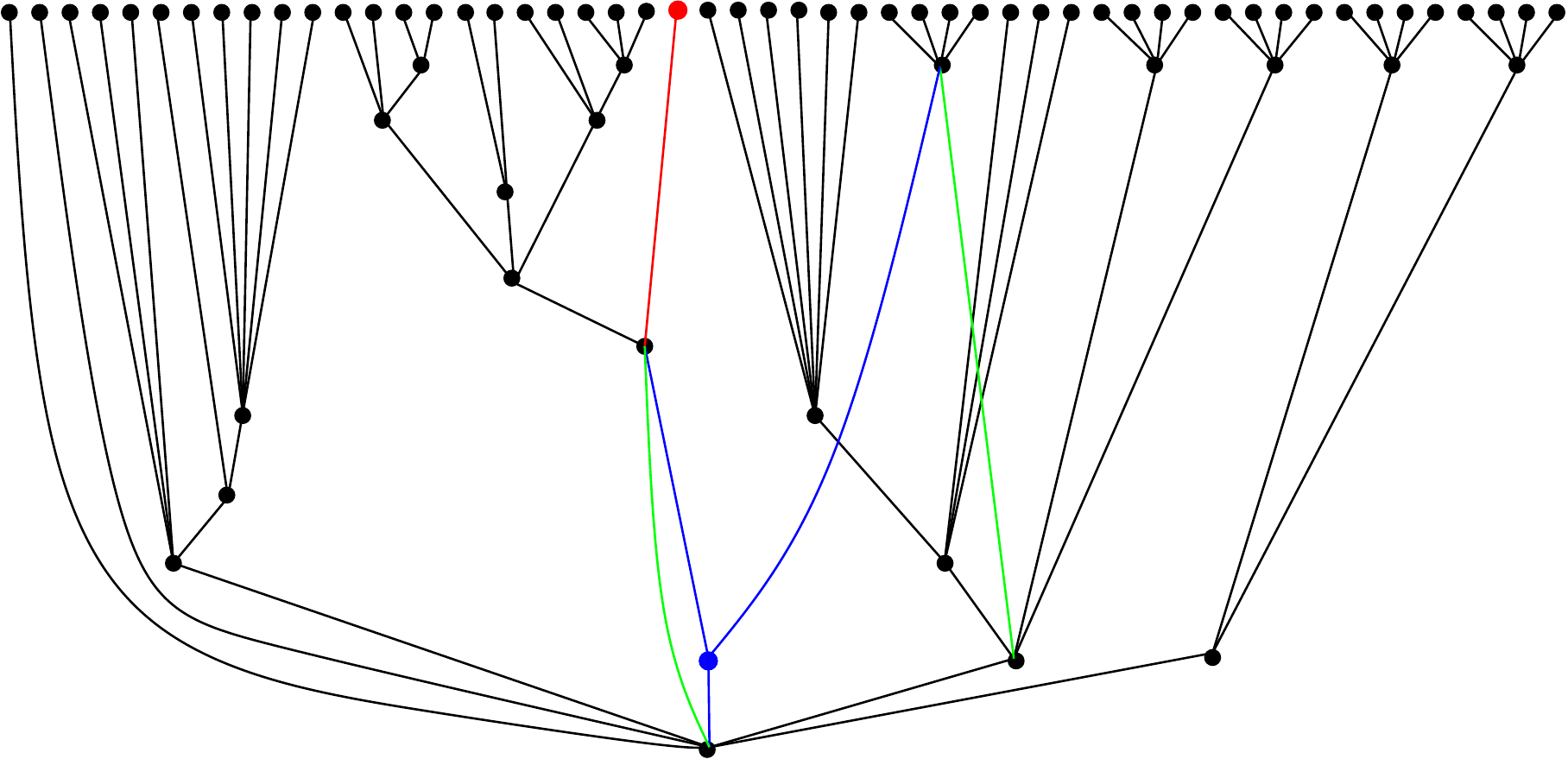}
\caption{Merging trees for the Morse potential with $N=9$ particles. The top row of the trees contain all SHS clusters and the trees follow unique Morse structures through the continuation process. The black nodes and edges are the same for each $\kappa$ value. The green nodes and edges are specific to the $\kappa_{\text{MED}}=49.5$ tree. The blue nodes and edges are specific to the $\kappa_{\text{LOW}}=23.4$ and $\kappa_{\text{HIGH}}=100.4$ trees, which are (interestingly) the same. The non-harmonic cluster and its path is shown in red. Note that only the topology of the trees is being shown, i.e. the vertical positions of nodes are not to scale.}
\label{N9Merge}
\end{figure}

\begin{figure}[!ht] 
\includegraphics[trim={0.25cm 0cm 0.5cm 0cm},clip,width=0.4\textwidth]{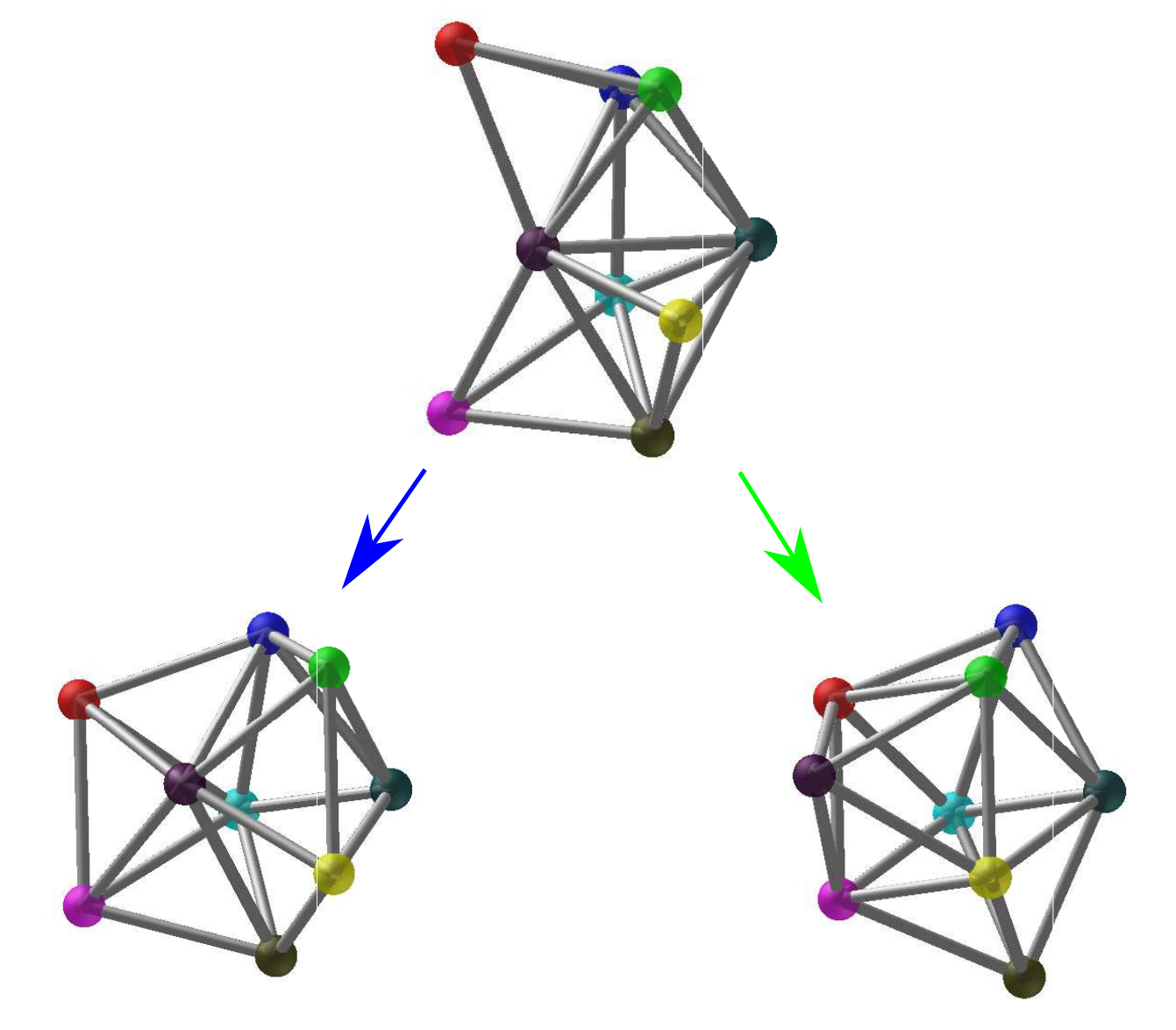}
\caption{The SHS cluster $10$ for $N=9$ (top), as well as the two possible results of the continuation procedure at $\rho=3$ for the Morse potential. Choosing $\kappa_{\text{MED}}=49.5$ results in the green path, while choosing  $\kappa_{\text{LOW}}=23.4$ and $\kappa_{\text{HIGH}}=100.4$ results in the blue path, consistent with the coloring in Figure \ref{N9Merge}. 
}
\label{N9diff}
\end{figure}

For $N=9$ the topology of the merging trees varies with both parameters and potential. The Morse trees are identical for short range, but not for long range  (Figure \ref{N9Merge}.)
The first difference occurs at $\rho=3$, where a single cluster merges with two distinct clusters depending on the value of $\kappa$. The SHS cluster as well as these two continuation possibilities are shown in Figure \ref{N9diff}. The SHS parent cluster is built from cluster 1 of $N=7$  by adding two extra particles, red and pink, along non-adjacent edges of the pentagon. The leftmost child cluster looks similar and can be formed by pulling the red and pink particles toward the center of mass, close enough to bond. The rightmost child cluster is quite different, containing an octahedron fragment the others do not. In addition to the red and pink particles being pulled closer together in this cluster, the dark purple particle seems to get pushed away from the center of the cluster. This child cluster is the only remaining cluster when the range becomes $1$, and has lower energy than the leftmost child cluster. We are not sure why the sticky parameter affects the result in this way or if there is physical intuition behind it. One possibility is that slight perturbations in the potential energy caused the optimization algorithm to find a deeper minimum.

For $N=10,11$, merging trees continue to depend on sticky parameter and potential, although the upper portions of the trees are still independent of parameters. For both potentials the trees are exactly the same for $\rho>32$ and $\rho>40$, respectively, though the differences for $\rho>30$, corresponding to a range of about $8$\% of particle diameter, in both cases are minimal; a difference of between $1$ to $10$ nodes. 

\begin{figure}[t]
\centering  
\subfigure[]{\includegraphics[width=0.45\linewidth]{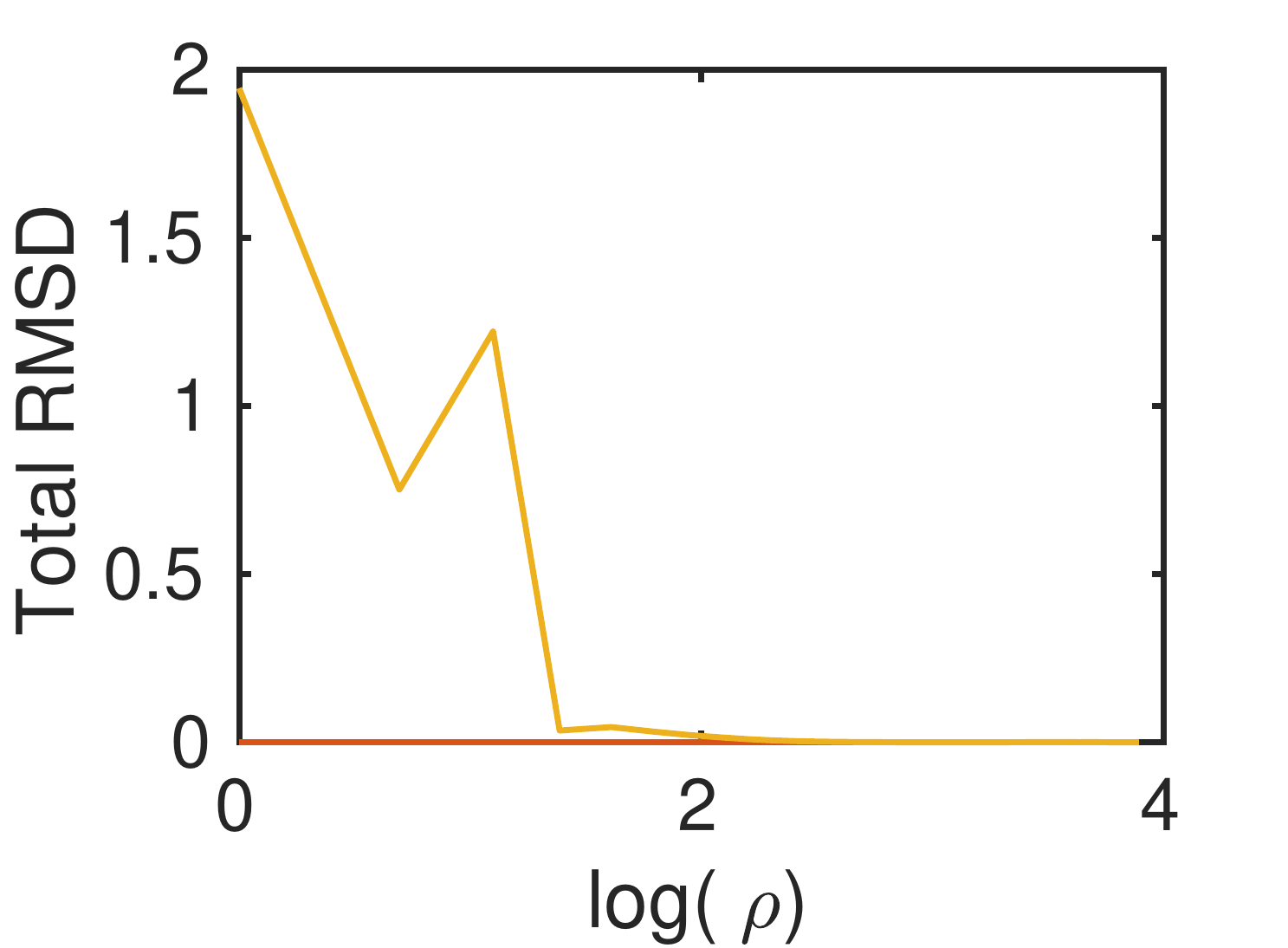}}
\subfigure[]{\includegraphics[width=0.45\linewidth]{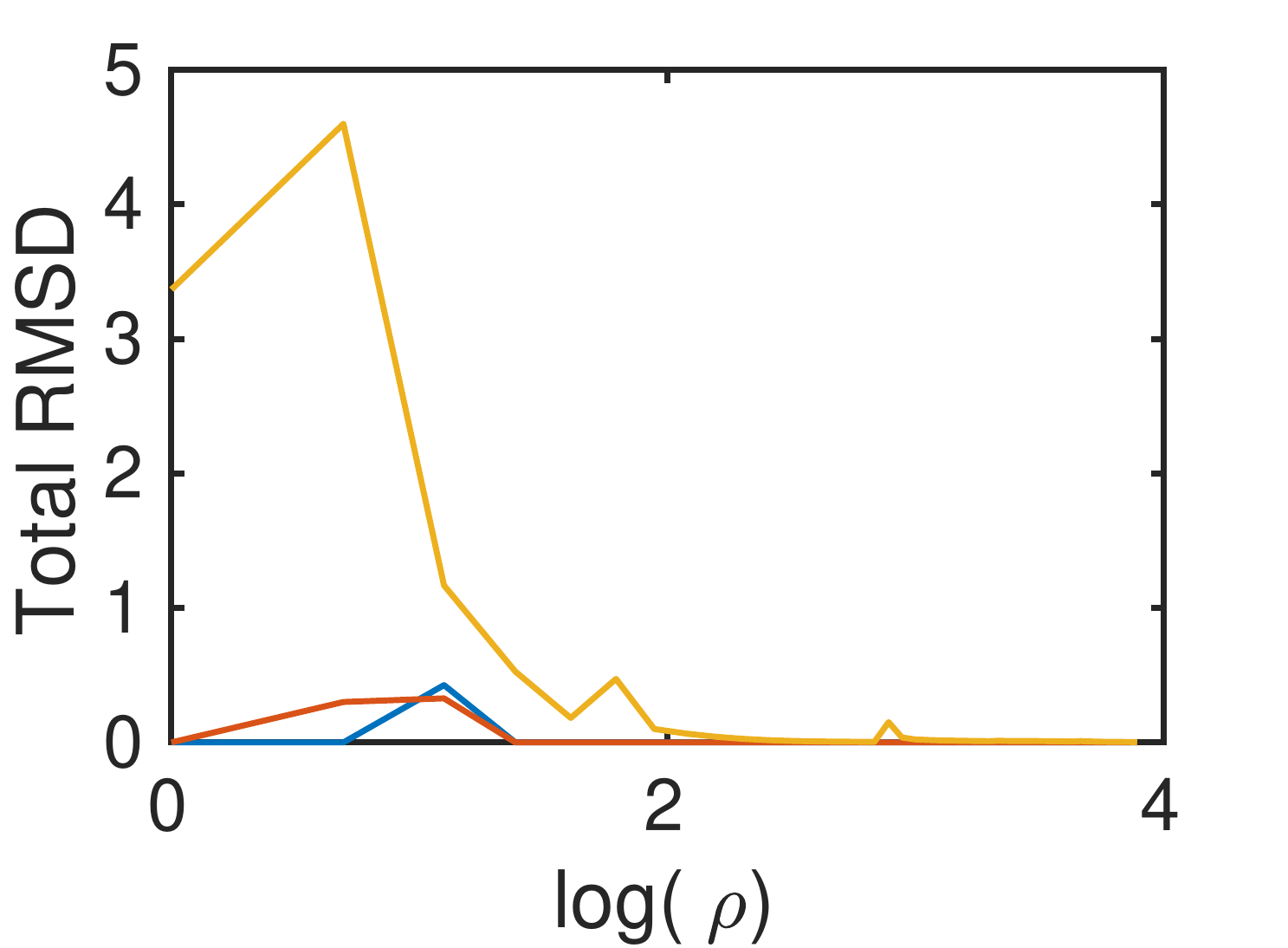}}
\subfigure[]{\includegraphics[width=0.45\linewidth]{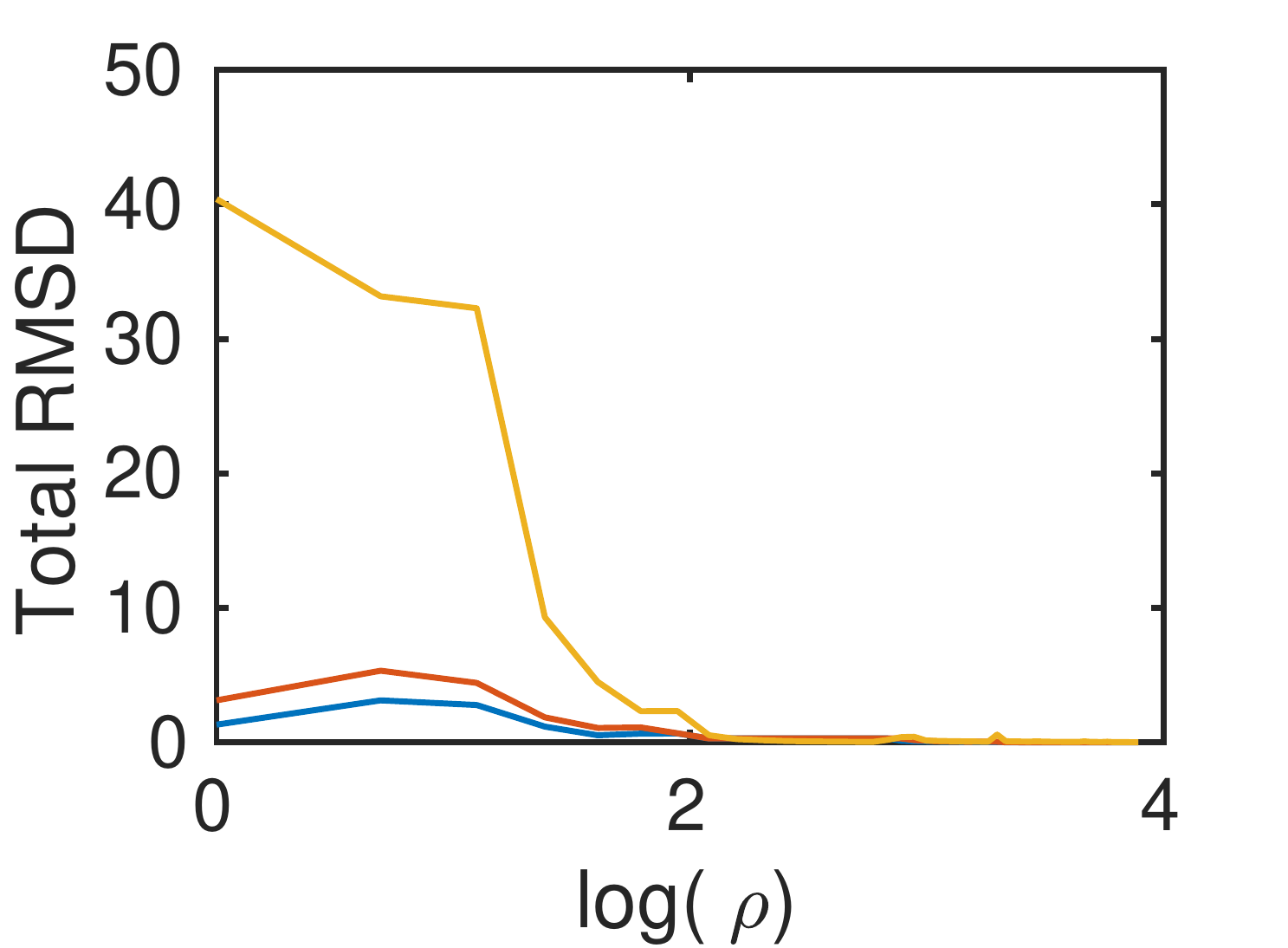}}
\subfigure[]{\includegraphics[width=0.45\linewidth]{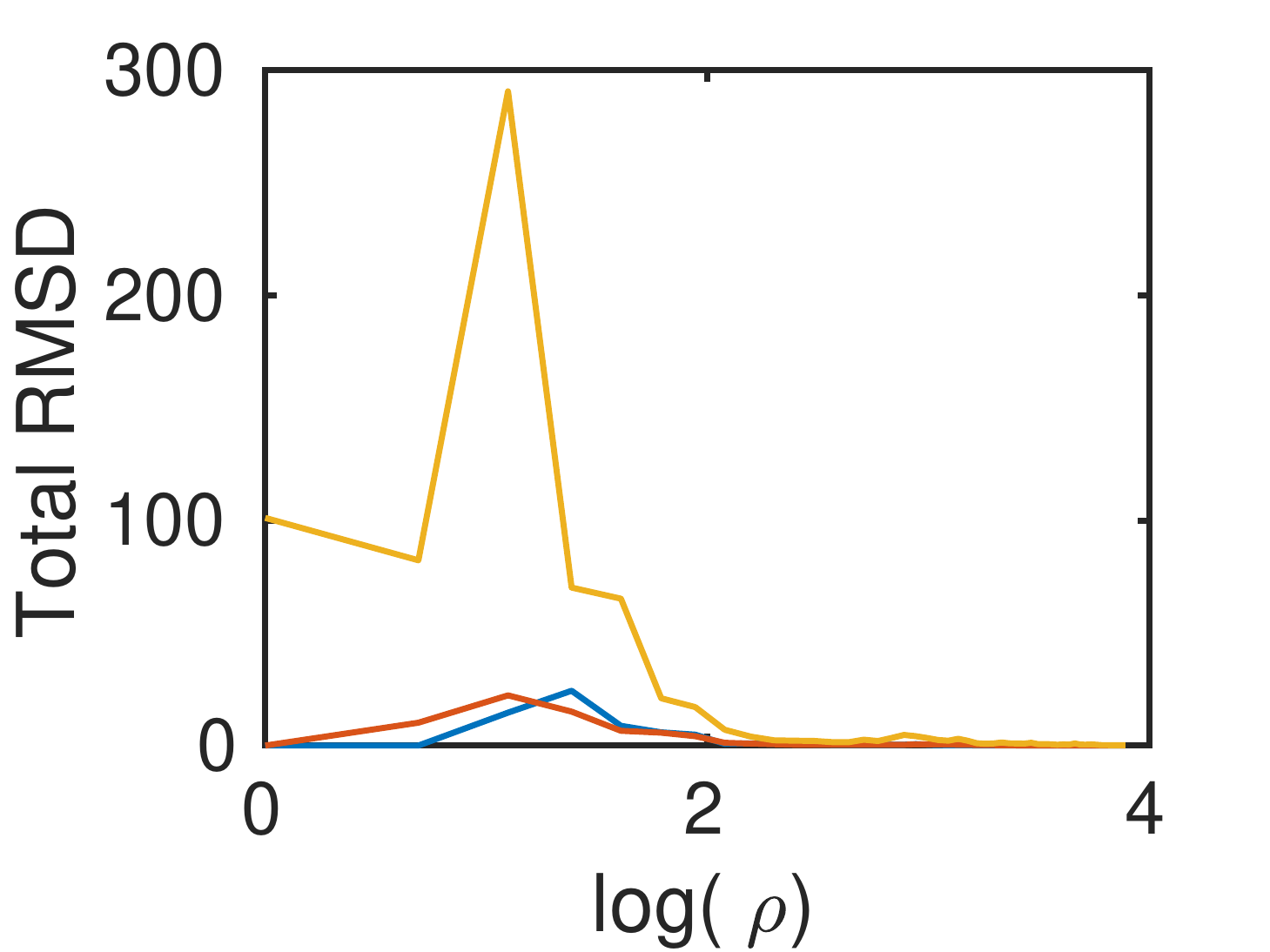}}
\caption{Total RMSD between clusters plotted as a function of $\log(\rho)$ for $N=8,9,10,11$ (a-d respectively). Blue curves compare low and medium values of $\kappa$ for the Morse potential, red curves compare low and medium values of $\kappa$ for the Lennard-Jones potential, and yellow curves compare Morse and Lennard Jones with medium $\kappa$ value.  To interpret the magnitude of these curves, note that a typical RMSD between a pair of distinct clusters is about $1-2$.  }
\label{editDistPlots}
\end{figure}

\begin{table}[t] 
\begin{tabular}{l | c c c | c c c | c c c}
 & \multicolumn{3}{c | }{Morse} & \multicolumn{3}{c | }{Lennard-Jones} & \multicolumn{3}{c}{Comparison} \\
 $N$  & L-M & M-H & L-H & L-M & M-H & L-H & L & M & H \\
\hline
6 	 & 0 & 0 & 0 & 0 & 0 & 0 & 0 & 0 & 0 \\
7     & 0 & 0 & 0 & 0 & 0 & 0 & 0.7 & 0.7 & 0.7  \\
8      & 0 & 0 & 0 & 0 & 0 & 0 & 4.1 & 4.1 & 4.1 \\
9  	& 0.6 & 0.6 & 0.1 & 0.6 & 0.1 & 0.6 & 12 & 11 & 12  \\
10    & 14 & 13 & 21 & 21 & 22 & 18 & 125 & 128 & 132 \\
11	 & 67 & 65 & 63 & 74 & 78 & 76 & 702 & 713 & 711 \\
\end{tabular}
\caption{Total RMSD summed over all integer values of $\rho$, for different comparisons. Left: Morse potentials compared at low (L), medium (M), and high (H) values of $\kappa$. Middle: Lennard-Jones potentials, also compared at different values of $\kappa$. Right: Morse and Lennard-Jones clusters compared at the same values of $\kappa$. 
}
\label{editDist}
\end{table}


For these larger values of $N$, it is hard to compare the topology of the trees quantitatively, so instead we compare the geometry of the individual clusters. Suppose we wish to compare two sets of clusters at a given range, where each set is obtained by performing continuation with different well-depth parameters $E$ or potential. There is a one-to-one mapping between the sets of clusters, because each cluster comes from a unique SHS cluster. We use this mapping to compare clusters: for each pair that comes from the same SHS cluster, we compute the root mean square deviation (RMSD; see Appendix \ref{testSame} for details.) We then sum the RMSD over all clusters to obtain a metric comparing the sets, which we call the total RMSD. We compute the total RMSD as a function of the range parameter $\rho$.

Figure \ref{editDistPlots} shows the total RMSD as a function of $\rho$ for $8\leq N\leq 11$, for clusters from different potentials or different $\kappa$ (different well-depths $E$, since the ranges at which they are compared is the same.) 
Remarkably, the total RMSD is nearly zero for all comparisons until relatively long ranges, roughly $\rho<6$, or about 35\% of particle diameter. This means that not only are the topologies of the merging trees nearly the same up to longer ranges, but, the geometries of the clusters themselves are also nearly the same -- in other words, for a fixed range, it doesn't matter whether you use a Morse or a Lennard-Jones potential, or what you choose for the well-depth (within the limits we considered); the metastable states are nearly identical. 

For longer ranges, the total RMSD increases the most for clusters from different potentials: the geometry of the metastable states is sensitive to the choice of potential. The total RMSD increases only a little bit for clusters from the same family of potential but with different well-depths; here the clusters have a much more similar geometry. The differences could be caused by slight differences in geometry, or by merging at slightly different ranges; we cannot tell the difference using this metric. 
Table \ref{editDist} shows a more extensive comparison than is contained in the figure, and supports the observation that there is more variety between families of potentials than within a single family.

\begin{figure}[t]
\includegraphics[width=0.42\textwidth]{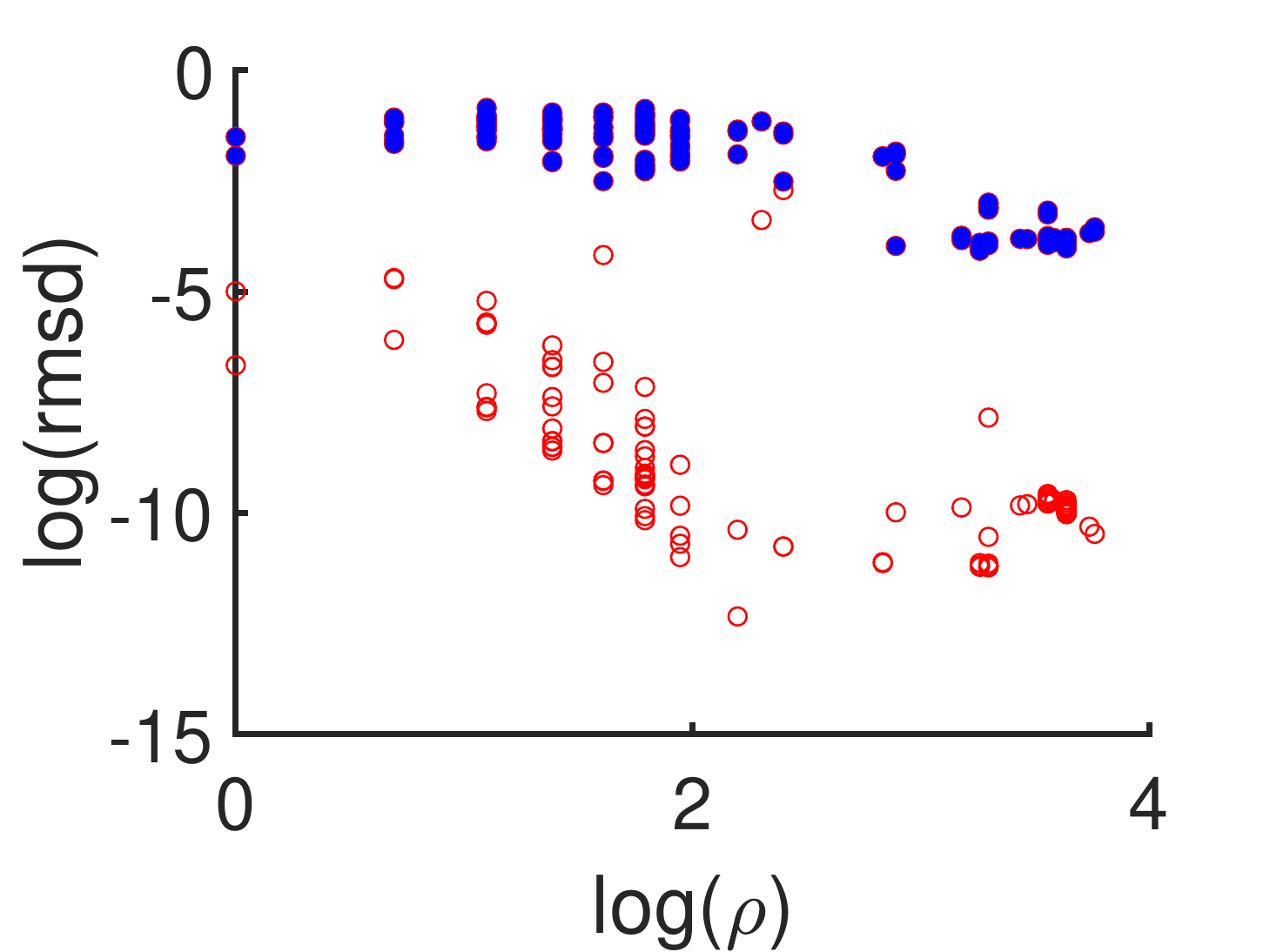} 
\centering
\caption{Scatter plot of RMSD values between a cluster before and after a merge at a given value of $\rho$, for all merge events for $N\leq 10$, with a log-log scale. For each group of merging clusters, the smallest RMSD is unfilled and the rest are filled in with blue. Each merge group has only one cluster with a small RMSD, with most being nearly $0$, implying that all merge events occur by fold bifurcations. 
}
\label{scatter}
\end{figure}

Finally, we return to the observation that for merges at small $N$, exactly one parent  transitioned smoothly, and we ask whether this holds true for larger $N$ as well. We consider every merge event for $N\leq 10$ and compute the RMSD between each cluster just before and just after it merged. For each group of merging clusters, we find exactly one cluster with a small RMSD, and all the others have much larger RMSDs (Figure \ref{scatter}.) Therefore, up to $N=10$ there is a unique smoothly-varying parent cluster for each merge event. This observation also suggests that the merges occur as fold bifurcations, in which a local maximum and local minimum annihilate, leaving no extrema. The annihilated local minimum then jumps abruptly in configuration space upon optimization past the bifurcation. The merges do not appear to occur as pitchfork bifurcations, in which two local minima separated by a local maximum smoothly coalesce into a single local minimum; such a bifurcation would give two smoothly-varying parents.

\subsection{Predicting merge events} \label{prediction}

\begin{figure}[!ht]
\begin{minipage}{0.23\textwidth}
    \subfigure[]{\includegraphics[trim={0.25cm 0cm 0.5cm 0cm},clip,width=\textwidth]{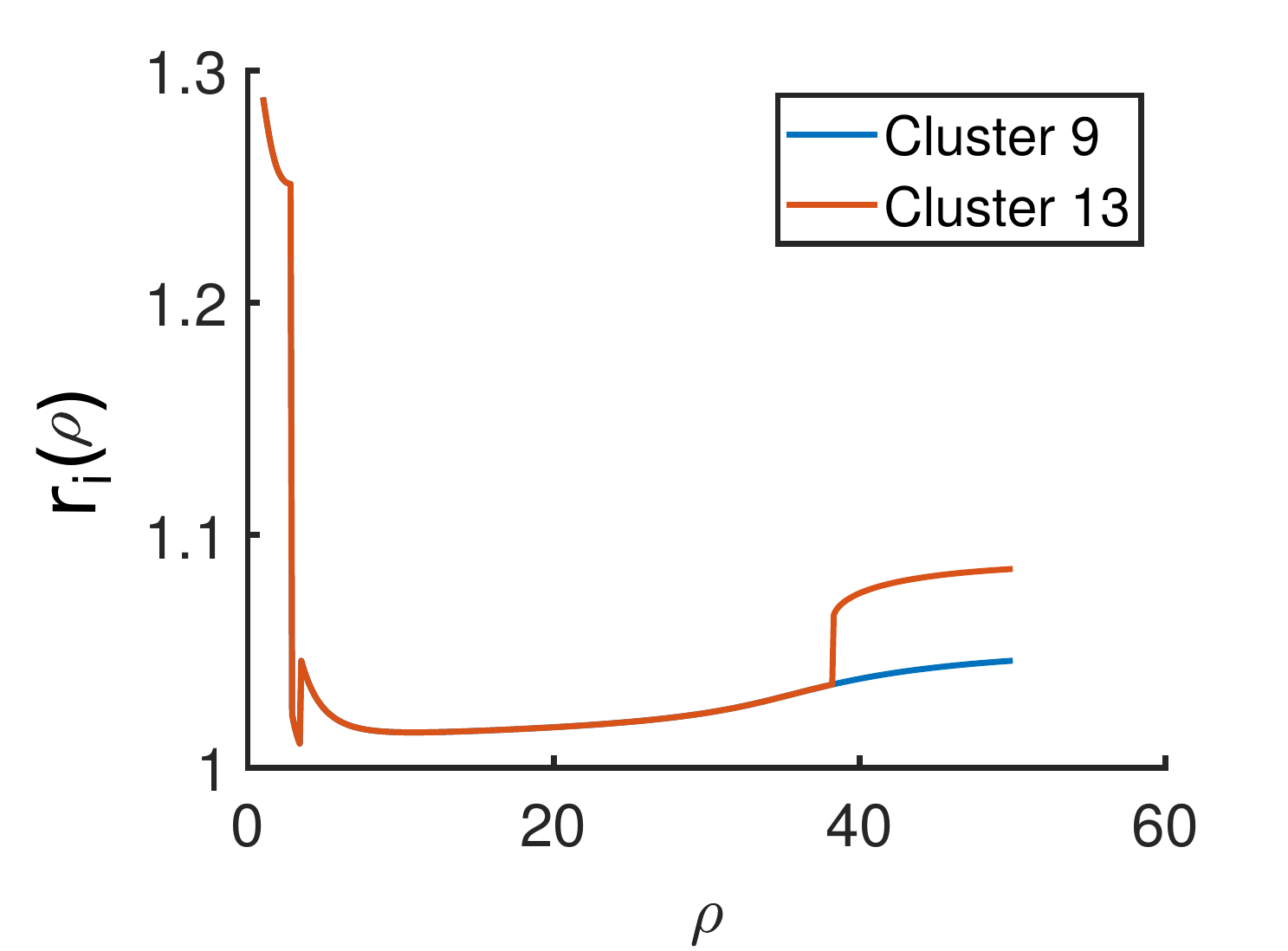}}
\end{minipage}
\begin{minipage}{0.23\textwidth}
    \subfigure[]{\includegraphics[trim={0.25cm 0cm 0.5cm 0cm},clip,width=0.6\textwidth]{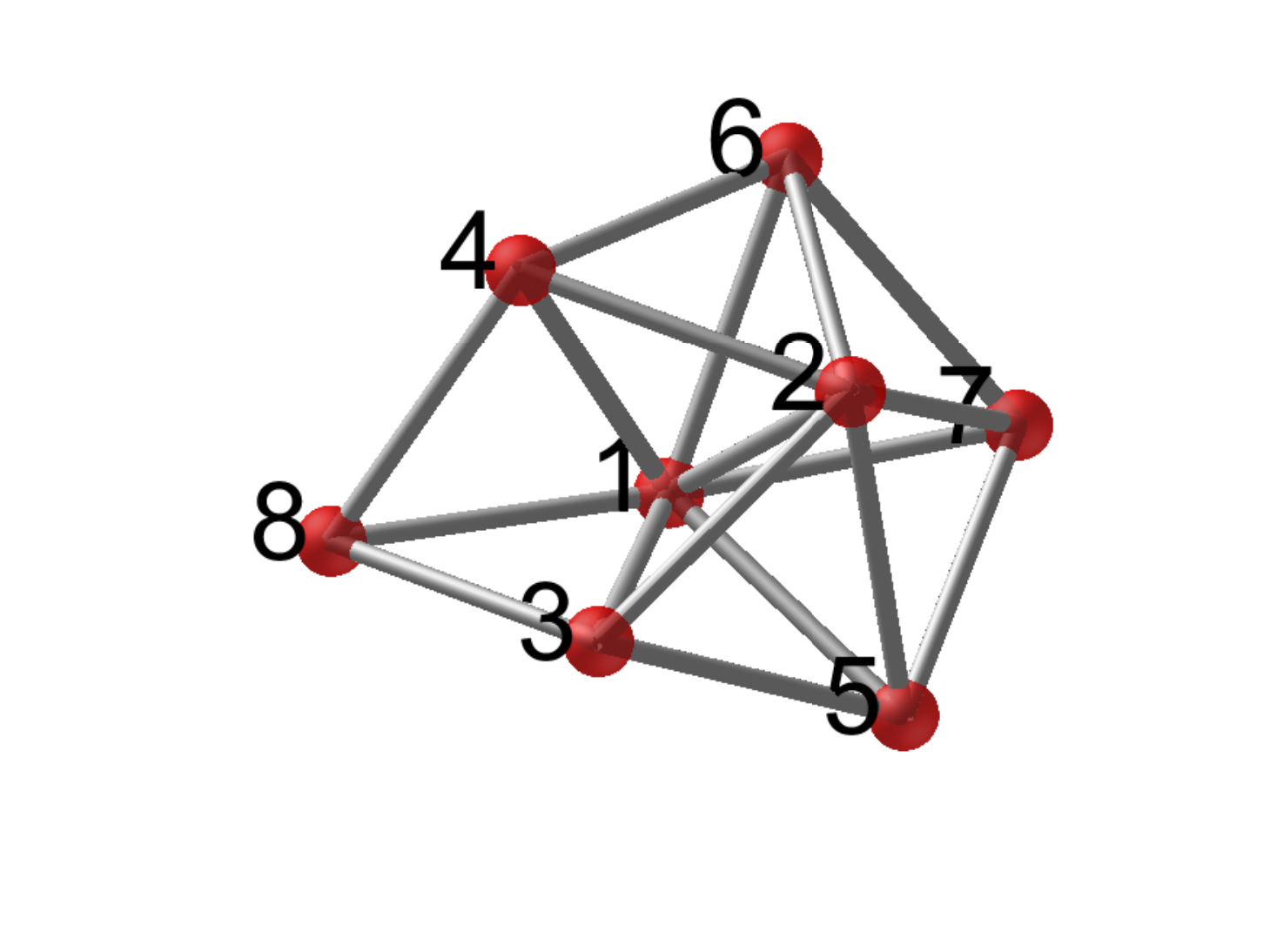}}
    \\
    \subfigure[]{\includegraphics[trim={0.25cm 0cm 0.5cm 0cm},clip,width=0.6\textwidth]{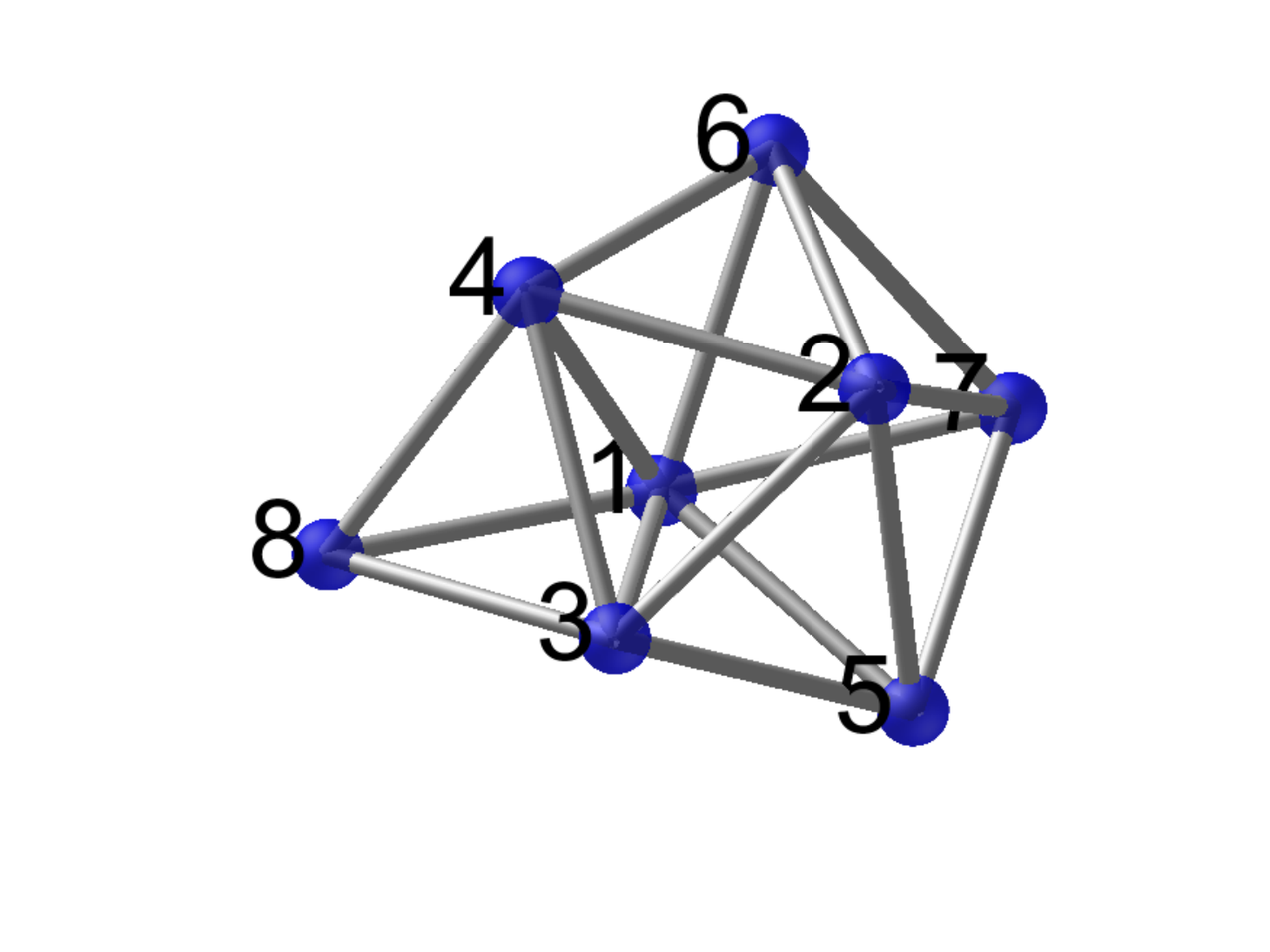}}
\end{minipage}
\caption{(a) $r_{13}(\rho),r_9(\rho)$, the minimum inter-particle distance greater than 1 for clusters 13 and 9 for $N=8$. Note that $r_9(\rho)$ varies smoothly but $r_{13}(\rho)$ jumps at $\rho=38$, when the clusters merge and become identical to cluster 9. 
The corresponding SHS clusters (13 and 9) are plotted in (b) and (c) respectively, in a way that minimizes the root mean square difference between them. The main difference between the clusters is the distance between particles $1$ and $2$ and particles $3$ and $4$. In cluster 9, $r_9(\rho)$ measures the distance between particles 1-2, and in cluster 13, $r_{13}(\rho)$ measures the distance between particles 3-4. These clusters merge  when particles $3$ and $4$ get close enough to bond. 
}
\label{N8rmin}
\end{figure}

\begin{figure}[!ht] 
\includegraphics[width=0.48\textwidth]{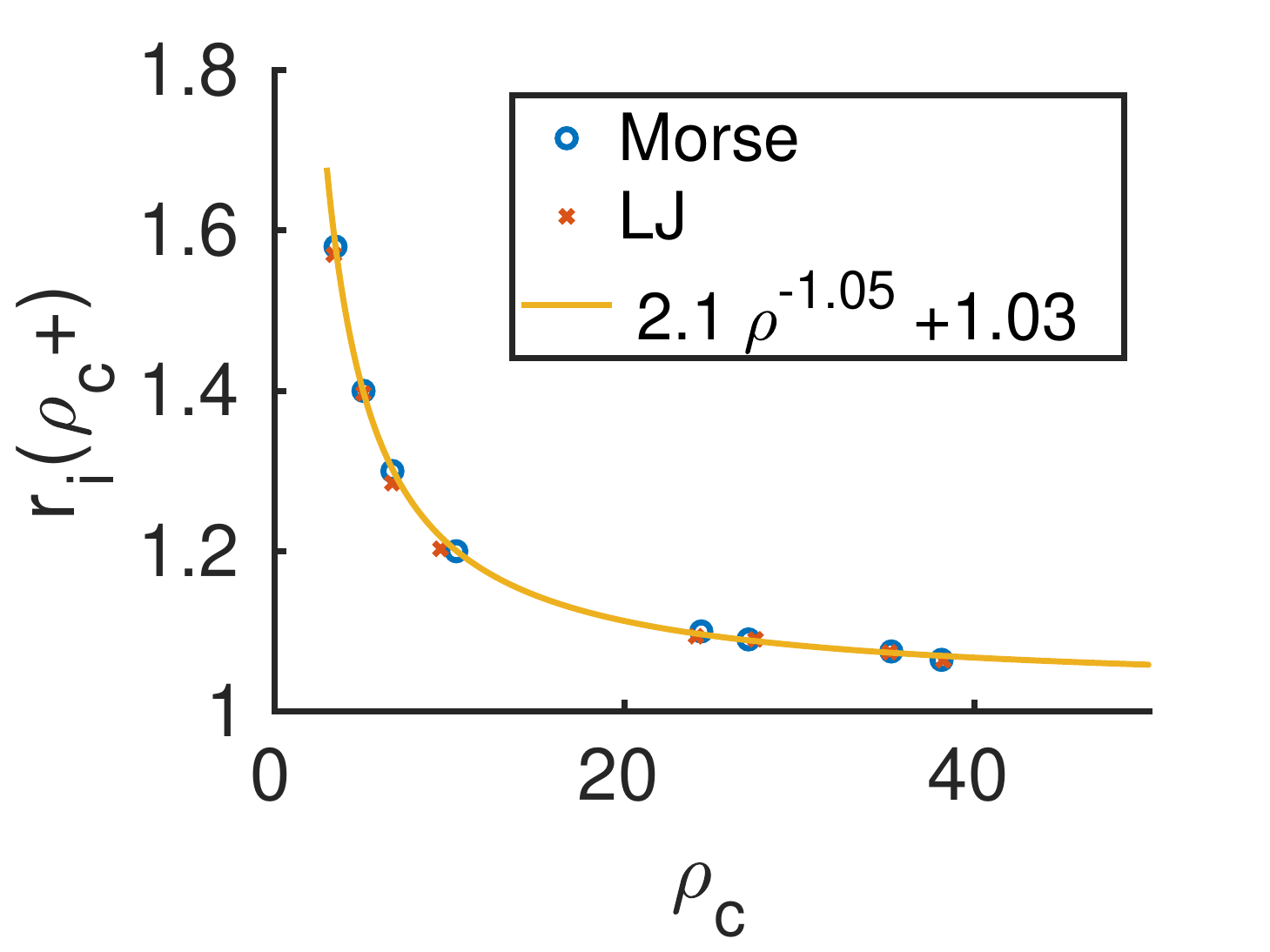}
\centering
\caption{The minimum interparticle distance greater than 1 right before a cluster rearranges, $r_i(\rho_c+)$, versus the value $\rho_c$ at which it rearranges for the first time, for both the Morse and Lennard-Jones potentials for $6\leq N\leq 10$. 
The scatter plot only includes merge events for clusters that rearrange discontinuously, and such that the parent SHS clusters were harmonic. 
}
\label{rMinRho}
\end{figure}

\begin{figure}[!ht] 
\includegraphics[width=0.48\textwidth]{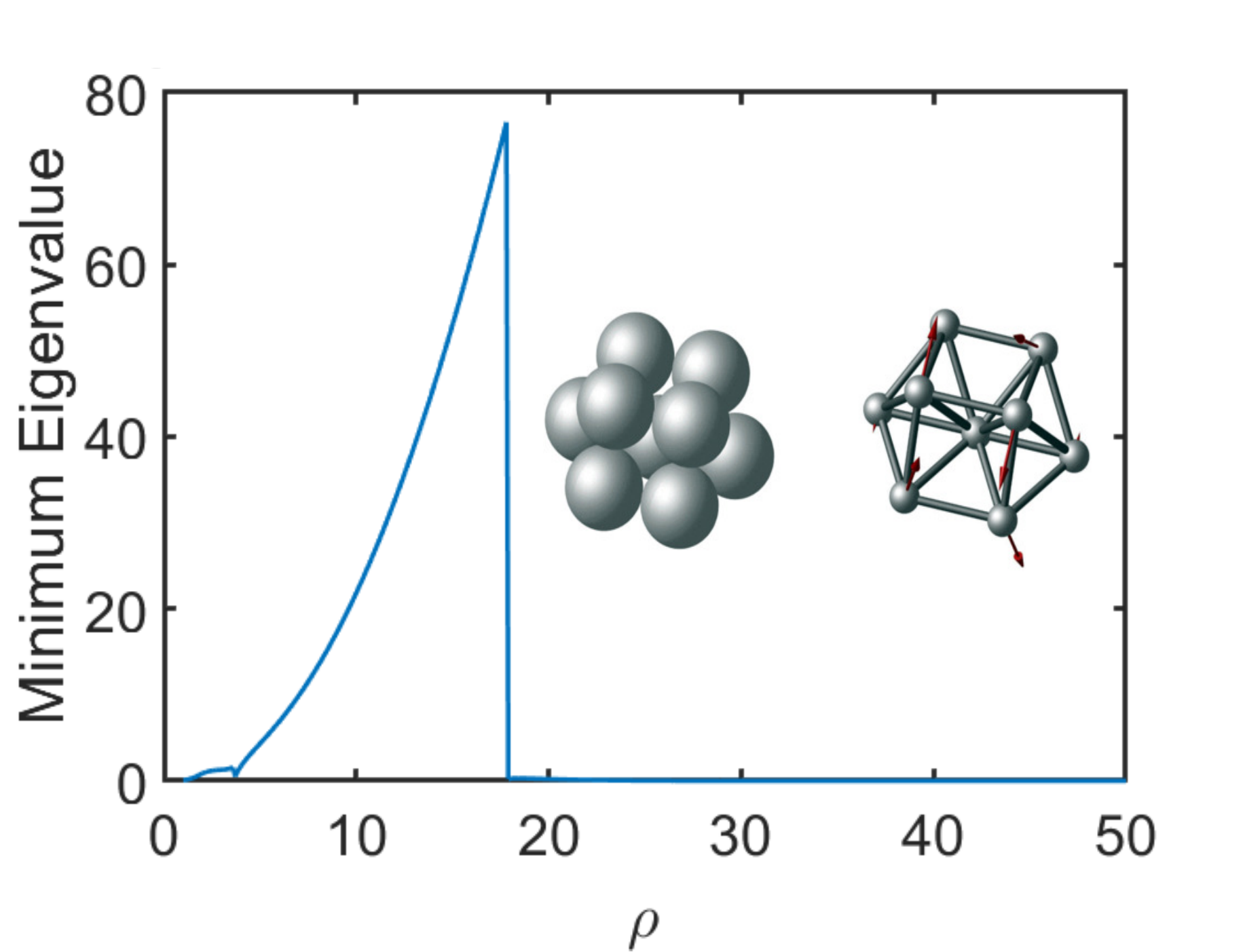}
\centering
\caption{Plot of the minimum eigenvalue of the Hessian of the Morse potential as a function of $\rho$ for the $N=9$ non-harmonic cluster. Plots of the SHS cluster taken from \cite{singularC} are shown as well: left plot shows particles with unit diameter, right plot has arrows on the sphere centers showing the zero mode of the Hessian.  
}
\label{singularEig}
\end{figure}

\begin{figure*}[!ht] 
\subfigure[]{\includegraphics[width=0.25\textwidth]{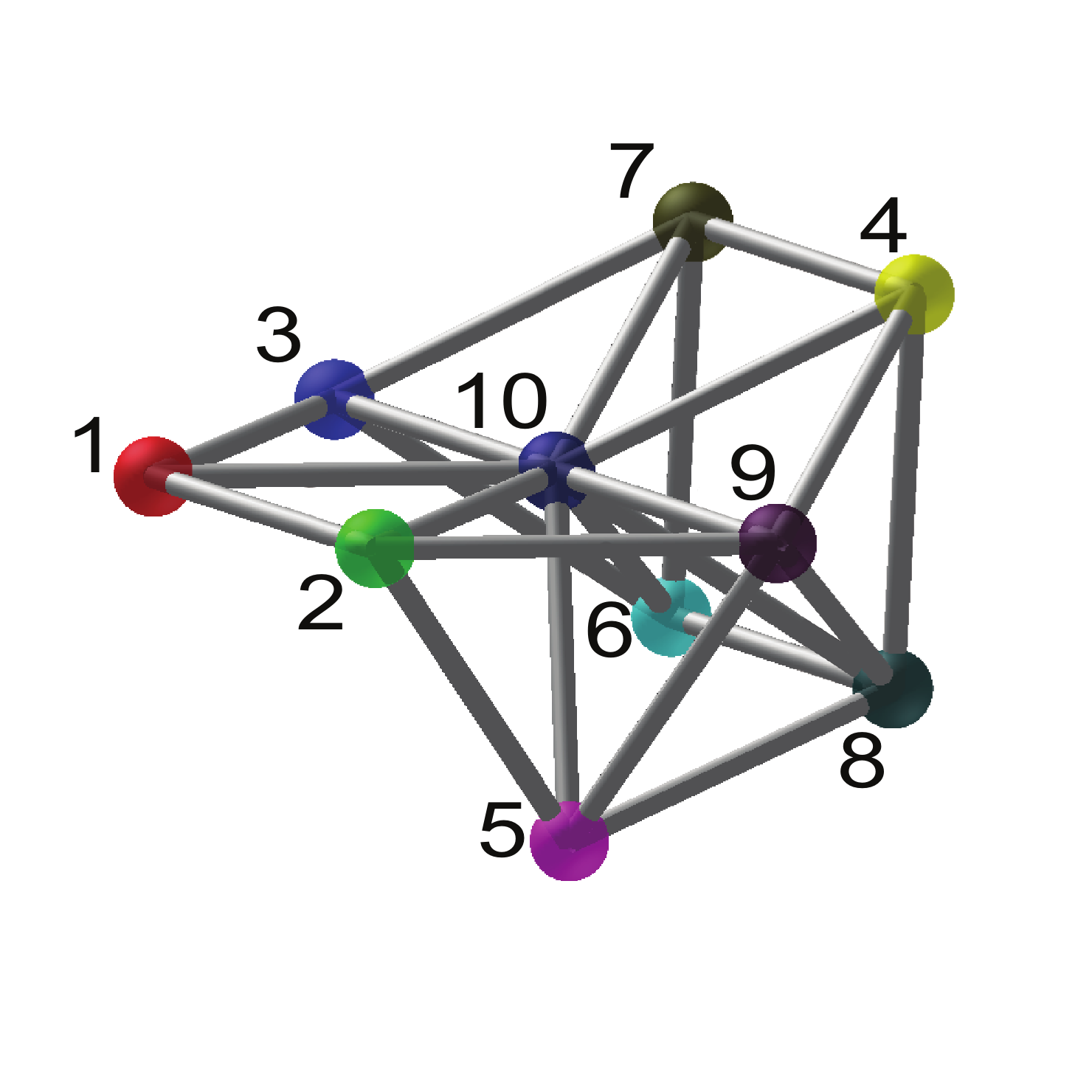}}
\subfigure[]{\includegraphics[width=0.32\textwidth]{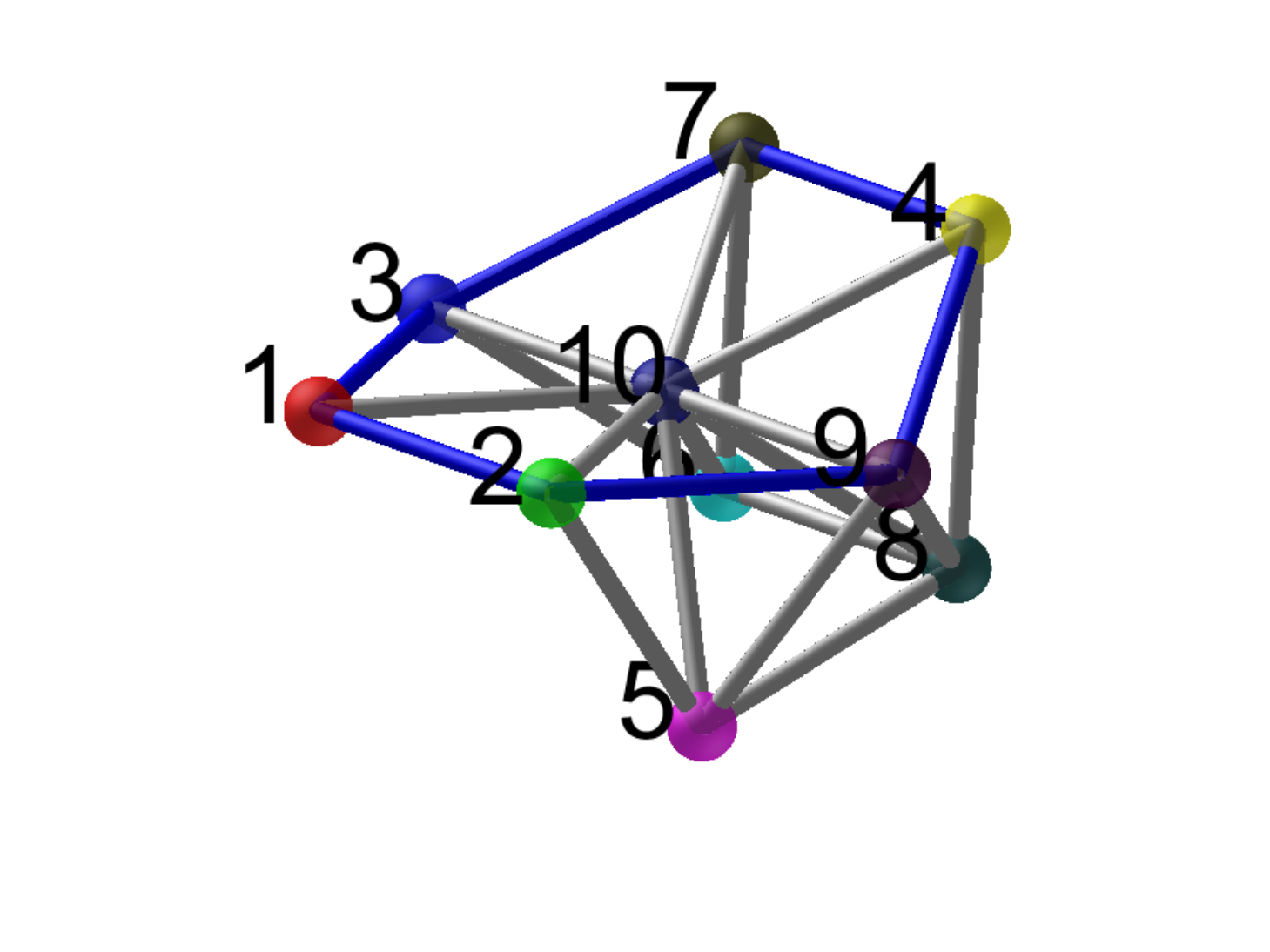}}
\subfigure[]{\includegraphics[width=0.39\textwidth]{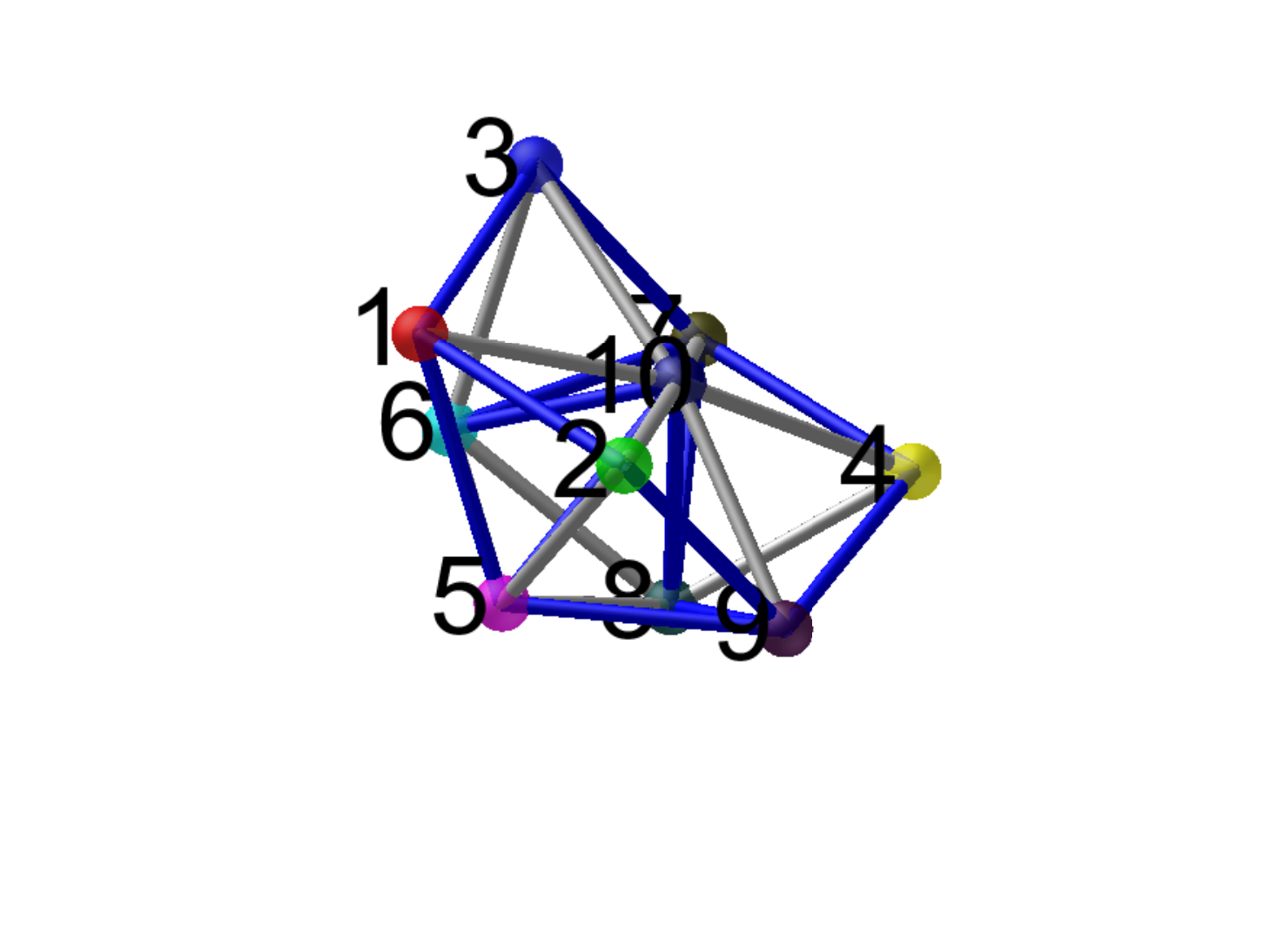}}
\caption{Evolution of cluster $6$ for $N=10$, a non-harmonic cluster, during the continuation process. (a) Starting SHS cluster. Note the near planar set of particles surrounding particle $10$. (b) Morse cluster at $\rho=35$. The bonds between the outer ring of particles have been pushed apart. (c) Morse cluster at $\rho=17$. The planar particles have been pulled down below particle $10$ now. At this point, the cluster has merged with another cluster that was not initially non-harmonic to start. Grey bars represent inter-particle distances less than or equal to $1$ and blue bars represent inter-particle distances less than $2.1\rho^{-1.05}+1.03$.
}
\label{singMerge}
\end{figure*}


Next we ask whether there is a geometric criterion that governs when clusters merge. 
Intuitively, we expect that a cluster rearranges when the range of the pair potential becomes comparable to the distance between a pair of non-bonded particles. 

To make this hypothesis quantitative, we define a function $r_i(\rho)$ to be the minimum inter-particle distance that is greater than 1, for cluster $i$ at range parameter $\rho$. Our hypothesis is that merges depend on  $r_i(\rho)$ in some way.  

To investigate how the function $r_i(\rho)$ behaves, we plot it as a function of $\rho$ in Figure \ref{N8rmin} for two Morse clusters at $N=8$, clusters $9$ and $13$ in \cite{clusters}. The SHS clusters have $r_9(\infty) = 1.0515, r_{13}(\infty) = 1.0887$.
As $\rho$ decreases, $r_9(\rho)$ decreases smoothly past $\rho=38$, whereas $r_{13}(\rho)$ jumps at $\rho=38$, from $r_i(38+) = 1.07$ to $r_i(38-) \approx 1.03$, when the cluster rearranges and becomes identical to cluster 9.

This behavior is the same for all merges that occur at $\rho=38$ for $7\leq N\leq 10$. In each merge, the cluster $i$ that rearranges has $r_i(\infty) = 1.0887$ and the cluster $j$ that varies smoothly has $r_j(\infty) = 1.0515$. Just before the merge, the values are also the same across all merges: $r_i(38+)\approx  1.07, r_j(38+)\approx 1.03$. 
Interestingly, the cluster that transitions smoothly is always the one that has the largest minimum eigenvalue in the Hessian at $\rho=50$.

One might guess from this data that $r_i(\infty)$, the minimum gap in the SHS cluster, directly determines the value $\rho_c$ at which a cluster $i$ first rearranges. Unfortunately, the story is not so simple: among clusters with $7\leq N\leq 10$, there are $101$ clusters with $r_i(\infty) = 1.0887$, with values of $\rho_c$ ranging from $31.60$ to $38.76$. To show this non-uniqueness, a scatter plot of $r_i(\infty)$ vs. $\rho_c$ is shown in Figure \ref{rMinRhoSHS} in Appendix \ref{shsMerge}. 

What is true is that $r_i(\rho_c+)$, the minimum distance for non-bonded particles just before the rearrangement, determines $\rho_c$. 
Figure \ref{rMinRho} shows a scatter plot of $\rho_c$ versus $r_i(\rho_c+)$ for all clusters $i$ that rearrange, for both potentials. 
The data is very well fit by the curve $r_i(\rho_c+) = 2.1\rho_c^{-1.05}+1.03$,  which we obtained using nonlinear least squares to fit the data with a function of the form $ax^b+c$. 
Since the width of the attractive well of the potential scales with $\rho$ as  $c\rho^{-1} + 1$, where $c$ is a constant, this fit is strong support for the hypothesis that a cluster's first rearrangement occurs when the closest non-contacting pair comes within the range of the pair potential. The distance of this non-contacting pair can change during the continuation, which is why the distance in the SHS cluster does not  determine the range at which the cluster rearranges (it does seem to determine it approximately, since the distance doesn't usually change too much during the continuation. See Appendix \ref{shsMerge}.) 
What is useful about our formula is that it gives a specific number with which to measure the width of the potential -- it tells us that when the gap between particles is closer than $2.1/\rho$, their interaction starts to matter. 

This formula does not hold for the smoothly-transitioning parents, which comprise about 1/3 of the clusters, and initially merge without rearranging. If we applied the formula to these clusters anyways, approximating $\rho_i(\rho_c+) \approx \rho_i(\infty)$, it would falsely predict a large $\rho_c$ (short range for rearrangement.) 
Some of these smoothly-transitioning parents do rearrange in later merges. 
We tried to find a relationship between $r_i(\rho)$ for these clusters, and the value of $\rho$ at which they first rearrange discontinuously, but we could not find any relationship.

Another exception to this behavior occurs for the nonharmonic clusters, which rearrange at much \emph{shorter} ranges than predicted by the formula above. Every non-harmonic SHS cluster for $9\leq N\leq 11$ has $r_i(\infty)\approx 1.4142$, so the formula above would predict they merge via re-arrangement at $\rho_c\approx 5$; using actual distances when they rearrange, which are closer to $r_i(\rho_c) \approx 1.3$, gives $\rho_c\approx 7$. However, most non-harmonic clusters undergo a large rearrangement at $\rho_c\approx 15-20$, well before the minimum gap is within the range of the potential. 

This suggests that nonharmonic clusters rearrange by a more global mechanism. To explore this mechanism, recall that nonharmonic clusters reach saddle points during the minimization for larger values of $\rho$ and a re-optimization procedure is performed. The result of the re-optimization is a cluster that is structurally very similar to the starting non-harmonic cluster, with a non-zero but very small minimum eigenvalue. The cluster stays close to this configuration until  $\rho\approx 15-20$, when it rearranges and merges with harmonic clusters. As an example, Figure \ref{singularEig}, which plots the minimum eigenvalue in the Hessian of the energy for the $N=9$ non-harmonic cluster. As $\rho$ decreases, the minimum eigenvalue slowly increases from $0$ until a jump occurs near $\rho=17$, at which point the cluster merges with a harmonic cluster. Similar behavior is exhibited for all $9\leq N\leq 11$ non-harmonic clusters except for 4 (of 35) non-harmonic clusters at $N=11$: they are nearly constant, with a small minimum eigenvalue, until they  rapidly rearrange at $\rho\approx 15-20$.  The exceptions at $N=11$ had a minimum eigenvalue that moved away from zero before the cluster merged.

 We examine the non-harmonic cluster $6$ for $N=10$ in detail.  This cluster, as well as most others, has a planar or near planar set of $6$ particles that attach to each other in a ring and to a seventh central particle. This cluster stays nearly the same until $\rho\approx 30$ when the particles on the outer ring begin to separate. This outer ring then begins to be pulled downward until $\rho\approx 17$ where the cluster rearranges and merges with another cluster. Various stages of this process are shown in Figure \ref{singMerge}. This general rearrangement mechanism occurred for most of the non-harmonic clusters. 

\begin{figure}[!ht] 
\subfigure[]{\includegraphics[width=0.17\textwidth]{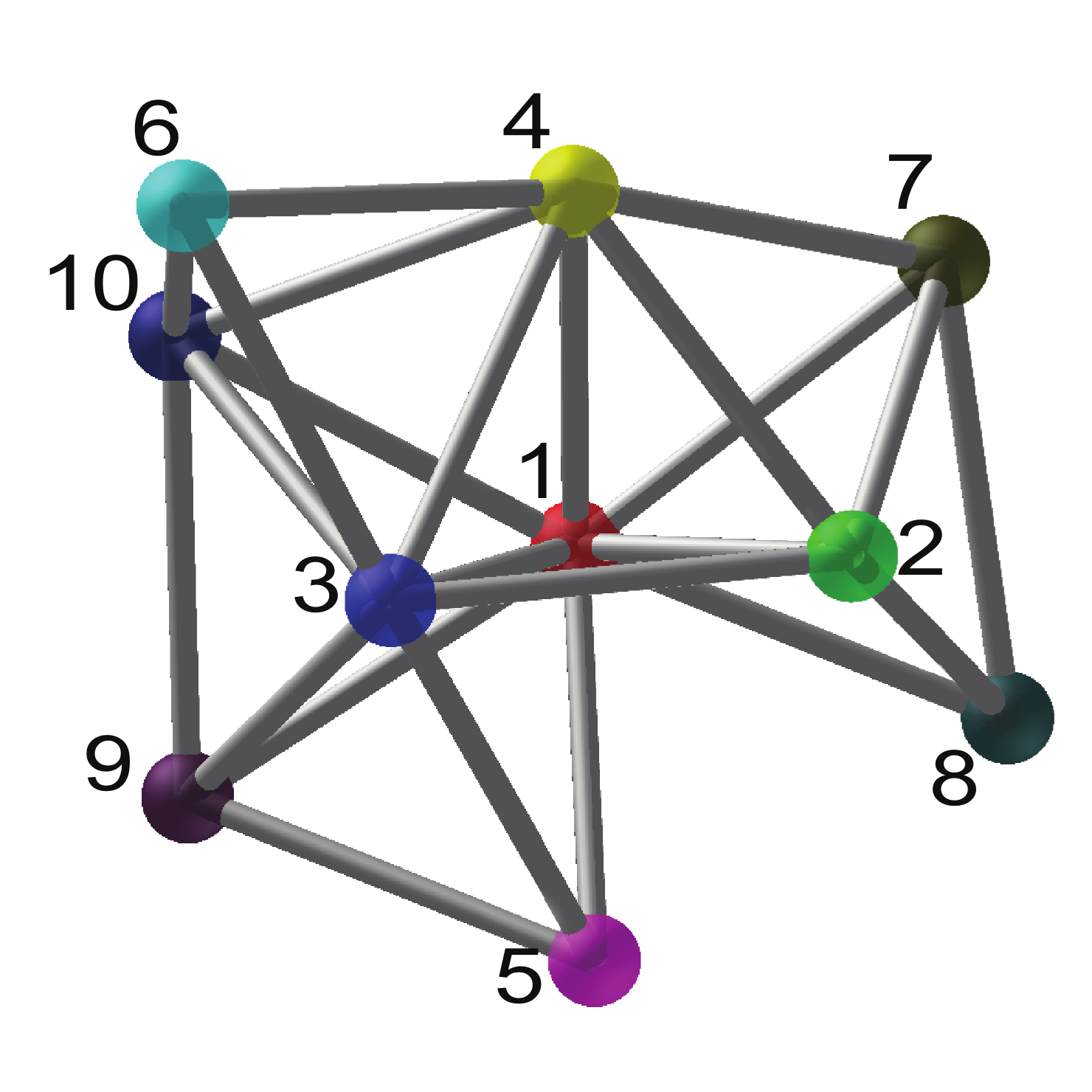}}
\subfigure[]{\includegraphics[width=0.17\textwidth]{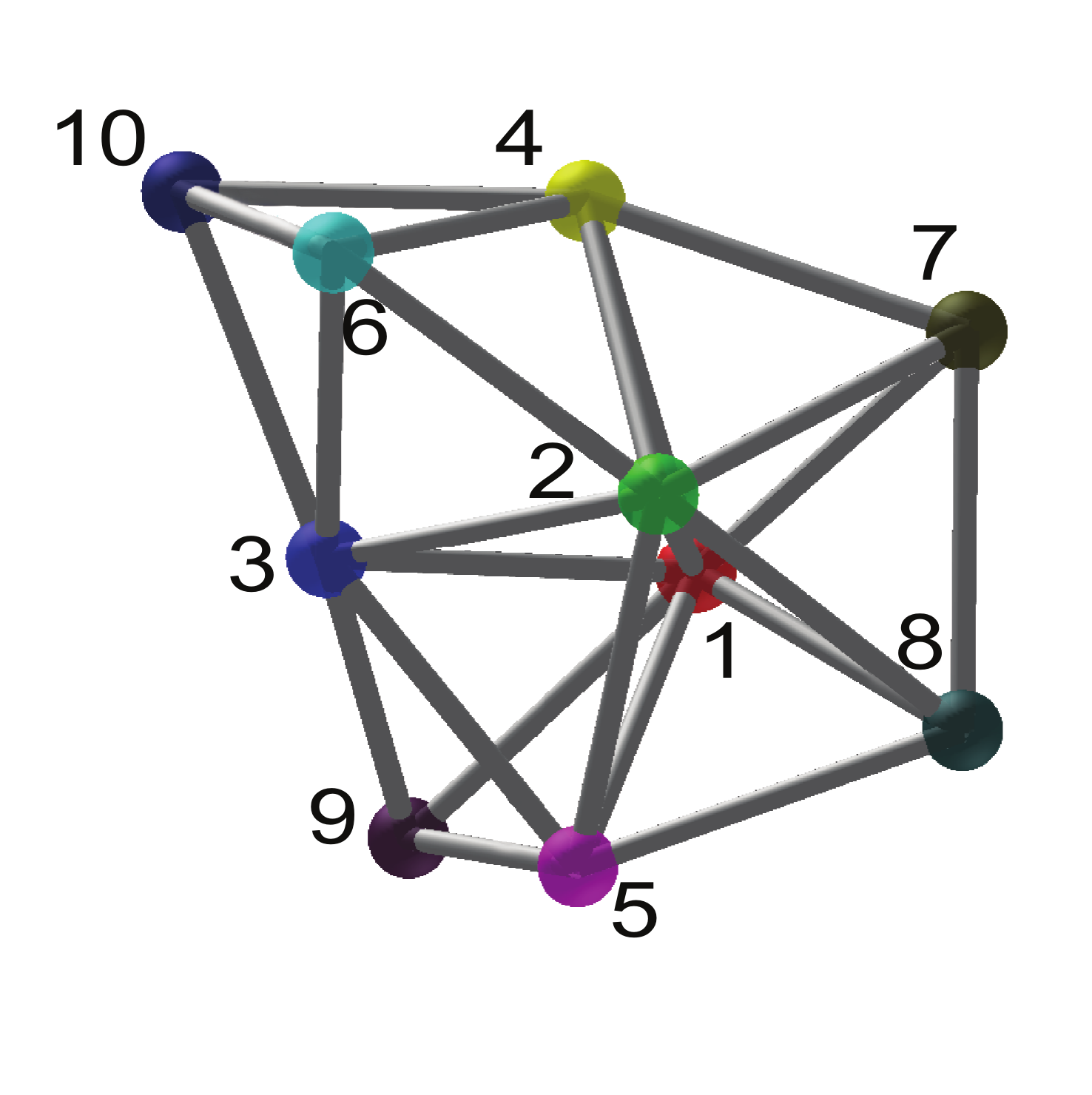}}
\centering
\caption{SHS clusters (a) $34$ and (b) $58$, for $N=10$. When a bar is added to cluster 34 between particles $5$ and $8$, the two closest non-bonded particles, it has the same adjacency matrix as cluster $58$. However, while these clusters both merge at $\rho=38$, they don't merge together. }
\label{counterEX}
\end{figure}

Despite being able to often predict \emph{when} a cluster will merge based on its geometry, we have not found a way to predict \emph{which clusters} will merge together. One idea was to compute an adjacency matrix for each cluster representing the pairs of particles whose distances are comparable to the minimum distance between non-bonded particles, and then to compare adjacency matrices. 
This doesn't work unfortunately; a counterexample is shown in Figure \ref{counterEX}. 


\section{Conclusion}

We used numerical continuation to study the evolution of sticky hard sphere clusters as the range of interaction increased, for Morse and Lennard-Jones potentials. 
This procedure finds most local minima of the smoother energy landscapes; the relatively few unmatched clusters were higher-energy clusters. 
This suggests that a similar technique could be used to find deep minima of larger systems, since SHS clusters with a maximal number of contacts can sometimes be found theoretically \cite{Bezdek:2012if,Bezdek:2018vt}, but exploring short-range energy landscapes numerically is a challenge because the potential develops very high gradients.

As the range of interaction increased, distinct clusters merged together, so the total number of unique clusters decreased. Interestingly, we found no non-trivial splitting events, where one cluster split into two unrelated by rotations or permutations. We analyzed the mechanism behind individual merge events in details for small clusters. Many larger clusters contain these smaller clusters as sub-units, so these simple merge events provide insight into how larger clusters merge. 

We found that all merges involved one cluster that varied smoothly, hence whose structure did not change, while the other clusters rearranged significantly during the merge. This suggests that all merges in our data were fold bifurcations. 
In addition, we found that the clusters that rearranged, did so when the range of the interaction potential became comparable to the minimum distance between particles that were not yet bonded. We found a specific formula to measure the relevant range: when the pairwise distance became about  $1+2.1/\rho$, these clusters rearranged. This formula may be useful for simulations or experiments that wish to use a threshhold distance for saying particles are bonded. 

An exception to this formula occurred for the non-harmonic clusters, which rearranged by a more global mechanism. 

We compared sets of clusters obtained by continuation for different potentials and different parameters, and found the corresponding clusters to be nearly identical for short and medium ranges, roughly 30\% of particle diameter, but varied quite a lot for longer ranges. There was more variation when we changed the family of potentials, than when we changed parameters within the same family of potentials. 




Our observations show that for short-ranged interactions, up to about $8$\% of particle diameter, the exact choice of pair potential and parameters have a negligible effect on the number of accessible ground states and the structure of local minima on the energy landscape; however, these states do differ from SHS clusters beyond a range of about $5$\% (for the values of $N$ considered.) For longer range potentials, greater than about $30$\% of particle diameter, the particular choice of interaction potential can affect the structure of these states. 
An intriguing extension would be to study the dynamics as the range increases, and to find a continuation procedure to study transition rates and paths between local minima or other dynamical quantities. Are they also similarly insensitive to the choice of potential? 




\section{Acknowledgements}

The authors would like to thank John Morgan for sharing data, Maria Cameron for providing code to determine point group order, and Dennis Shasha for helpful discussions.
This work was supported by the United States Department of Energy under Award no. DE-SC0012296, and the Research Training Group in Modeling and Simulation funded by the National Science 
Foundation via grant RTG/DMS – 1646339. 
M. H.-C. acknowledges support from the Alfred P. Sloan Foundation.

\bibliographystyle{unsrt}
\bibliography{references}{}

\section{Appendix}
\subsection{Optimization Algorithm \label{CG}}
The conjugate gradient algorithm \cite{numRecipes} was used to minimize the potential energy. For $N$ particles, the potential energy is a function of $3N-6$ position variables. There are six degrees of freedom corresponding to rigid body translation and rotations, which are removed by constraining particle $1$ to the origin, particle $2$ to the $x$-axis, and particle $3$ to the $x-y$ plane. The conjugate gradient algorithm terminates when the change in the objective function is less than $10^{-13}$ and it is verified that the point is a local minimum via the eigenvalues of the Hessian. The condition number of the minimization problem scales with $E$, so the convergence rate scales like $1-\frac{1}{\sqrt{E}}$ for $E\gg 1$. However, for large enough $E$, $E\sim\log\kappa$, so the convergence rate does not change much with sticky parameter. This is consistent with what we see in practice. 

The conjugate gradient method is unstable from some starting points and can  blow up or take very large steps. This depends on the value of the sticky parameter, as a larger sticky parameter corresponds to a deeper well-depth, which results in steeper gradients. A check for instability is performed after every optimization, and if either possibility occurs, we reset and try one of a variety of methods. The methods are, in order of application, gradient descent followed by conjugate gradients, conjugate gradients with resets every $3N-6$ iterations, swapping two random particle labels among the last $N-3$ particles and then applying conjugate gradients, or perturbing the starting point by a random vector of norm $10^{-12}$ and applying conjugate gradients. If all of these methods fail, the starting point is logged as the minimum. This usually results in a point with potential gradient norm $10^{-7}E$, instead of the usual tolerance of $10^{-13}E$. The fraction of optimizations that result in such an error are $\left( 5.6,3.5,1.5,0.74,0.6\right) \%$ for $6\leq N\leq 10$. 

When a saddle point is reached (minimum eigenvalue becomes negative), a re-optimization procedure is applied to reach a local minimum. This involves displacing along the eigenvector corresponding to the negative eigenvalue and re-applying conjugate gradients until a minimum is found. In some cases, a saddle point with more than one negative eigenvalue is reached; usually only $2$, but occasionally more. We found that the choice of eigenvector to displace along did not affect the result of re-optimization, so the eigenvector corresponding to the eigenvalue of largest magnitude is chosen for consistency. 

A C++ implementation of this continuation procedure using dlib \cite{dlib09} and additional code used (merge detection, RMSD, etc.) is available on GitHub \cite{shs_github}.

\subsection{Testing Uniqueness\label{testSame}}
To determine when two clusters are the same, we begin by checking that they have the same set of inter-particle distances. If so, the Kabsch algorithm \cite{Kabsch} is applied to compute an optimal rotation of one cluster onto the other, and the root mean square deviation is computed as
$$\text{RMSD}=\left(\frac{1}{N}\sum_{i=1}^N D_i^2\right)^{0.5},$$
where $D_i$ is the Euclidean distance between particle $i$ in cluster one and cluster two. If this is less than a tolerance of $10^{-6}$, we consider the clusters the same. 

If not, we check possible permutations. To do so efficiently, we begin by computing the moment of inertia tensor, diagonalizing it, and using it to rotate each set of particles to the principal axes. Then we need to re-align (permute) the particles so both clusters have the same configuration. We do this by solving a minimum cost assignment problem, where the cost matrix is the Euclidean distance between particle $i$ in cluster one and particle $j$ in cluster two. We then re-apply the first part of the algorithm. 

This still does not account for certain reflectional and rotational symmetries, so we repeat this procedure for all swaps and sign combinations of $(x,y,z)$ coordinates, $48$ possibilities. Taking the minimum value among all these possibilities gives the RMSD. 

In the case of $12$ particles, this procedure becomes too slow to compare all pairs of clusters. In this case, we adopt a heuristic approach where we simply compare a sorted list of inter-particle distances between clusters. This is a necessary condition for uniqueness but may not be sufficient. Up to $N=11$, the heuristic approach gives the same results as comparing RMSD values, so we are hopeful this extends to $N=12$. 

\subsection{Sticky Parameter\label{stickyDeriv}}

Here we discuss the origin of the sticky parameter, $\kappa$, and its relation to the second virial coefficient, $B_2$. In our numerical experiments, we seek to vary the range, $\rho$, of potentials, $U(r)$, in our continuation procedure. This leaves a free parameter in the potential function, $E$, the well-depth. Keeping $E$ fixed as the range varies changes properties of the potential function at every stage. One such property is how ``strong" a bond between two nearby particles is, which is given by the partition function 
\begin{equation} 
Z = \int_0^{r_c} e^{-\beta U(r)}dr,
\end{equation}
where $r_c$ is a cutoff beyond which the pair potential is essentially constant. 
Thus a natural way to choose $E$ after changing the range is to pick $E$ such that $Z=\kappa$ remains constant. 
To approximate this integral, we can use the method of Laplace asymptotics. The potential function has a minimum at $r=d$, such that $U'(d)=0$, and $U''(d) > 0$. We expand the potential function in a second order Taylor series to get $U(r)\approx U(d) +\frac{1}{2}U''(d) (r-d)^2$ near the minimum. The integrand then becomes a Gaussian, and since it decays very fast away from the minimum, we may extend the limits of integration to $\pm\infty$. Evaluating the Gaussian integral gives us the expression given for $\kappa$ in the introduction,
\begin{equation}
    \kappa \approx e^{-\beta U(d)} \int_{-\infty}^\infty e^{ -\frac{1}{2}\beta U''(d) (r-d)^2} dr = \sqrt{\frac{2\pi}{\beta U''(d)}}e^{-\beta U(d)}.
\end{equation}

We can show that our sticky parameter is related to the more commonly used  second virial coefficient, $B_2$. The second virial coefficient can be expressed as \cite{tuckerman_SM}
\begin{equation}
    B_2(\beta) = -\frac{1}{2} \int_0^\infty \left( e^{-\beta U(r)} - 1\right) 4\pi r^2 dr,
\end{equation}
and appears as the second coefficient in a power series correction to the ideal gas law for a given interaction potential. Again, we can approximate $B_2$ by using Laplace asymptotics in the same way. We first note that for $r\gg d$, $U(r)\approx 0$. Therefore the term $\left( e^{-\beta U(r)} - 1\right)$ goes to $0$ exponentially fast, contributing nothing to the integral. Near the minimum, $\left( e^{-\beta U(r)} - 1\right) \approx e^{-\beta U(r)}$, and we can proceed as before. We make a second order approximation of the potential and evaluate the second moment of the Gaussian integral. The final result is
\begin{align*}
    B_2(\beta) &\approx -\frac{1}{2} e^{-\beta U(d)} \int_{-\infty}^\infty 4\pi r^2 e^{-\frac{1}{2}\beta U''(d)(r-d)^2} dr\\
    &= -\left(\frac{2\pi}{\beta U''(d)}\right) ^{3/2} e^{-\beta U(d)} \\
    &= - \frac{2\pi}{\beta U''(d)} \kappa,
\end{align*}
showing $B_2$ is a approximately linear function of the sticky parameter. 

\subsection{Predicting Merge Events from SHS Geometry} \label{shsMerge}

\begin{figure}[t] 
\includegraphics[width=0.48\textwidth]{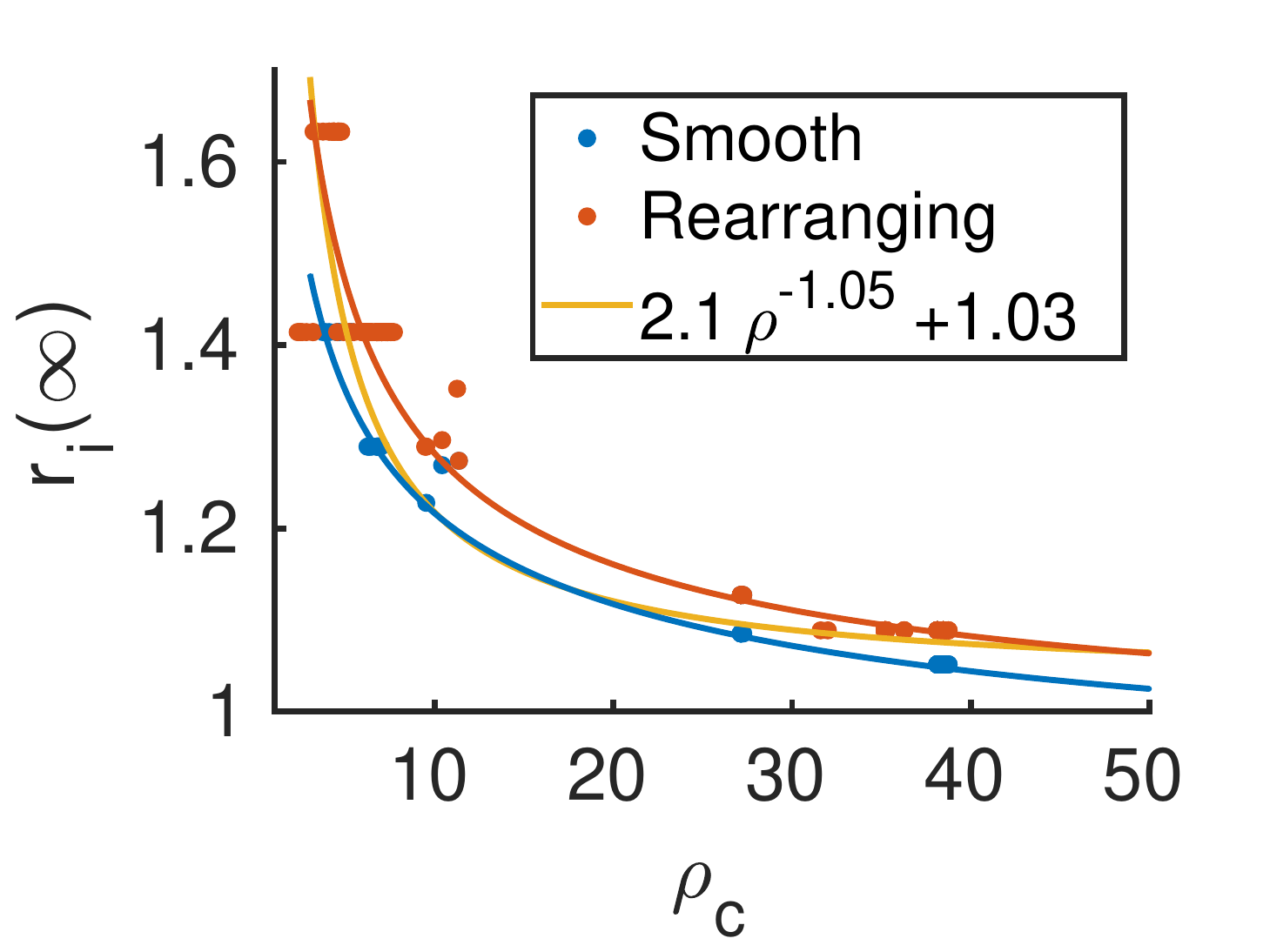}
\centering
\caption{Scatter plot of the minimum distance between non-bonded particles in a SHS cluster, $r_i(\infty)$, vs. the range value where the cluster merges, $\rho_c$, for all harmonic clusters with $6\leq N\leq 10$. Blue points show clusters which merge smoothly, red points show clusters which merge by rearranging. The blue curve is fit by $1.03\rho_c^{-0.43}+0.83$, and the red curve is fit by $1.45\rho_c^{-0.63}+0.94$. The yellow curve shows the fit for $r_i(\rho_c+)$, which sits between the red and blue curves. 
}
\label{rMinRhoSHS}
\end{figure}

We established a relationship between the value of the range parameter in which a cluster rearranges, $\rho_c$, and the minimum distance between non-bonded particles just before this rearrangement, $r_i(\rho_c+)$. Ideally, a one-to-one map between $\rho_c$ and $r_i(\infty)$, coming from the Sticky Hard Sphere geometry, would allow us to predict when any structure would merge. Figure \ref{rMinRhoSHS} shows a scatter plot of $r_i(\infty)$ vs. $\rho_c$, with smooth transitioning clusters in blue and rearranging clusters in red. The curves are fit using non-linear least squares with the form $ax^b+c$. 

Unfortunately, we find that the map is not one-to-one. There are many points for which $r_i(\infty)=1.0887$ and $r_i(\infty)=\sqrt{2}$, but the corresponding $\rho_c$ values are spread over a large range. Despite this, $r_i(\infty)$ does provide us with an approximation of when clusters will merge. This minimum distance can evolve differently as a function of cluster geometry, but the variations are typically small. Hence, if $r_i(\infty)=1.0887$ or $r_i(\infty)=\sqrt{2}$, you can safely state the cluster will merge \emph{around} $\rho=38$ and $\rho=6$, respectively. 

\subsection{$N=8$ Merging Trees}\label{n8trees}

\begin{figure*}[t]
\includegraphics[width=1\linewidth]{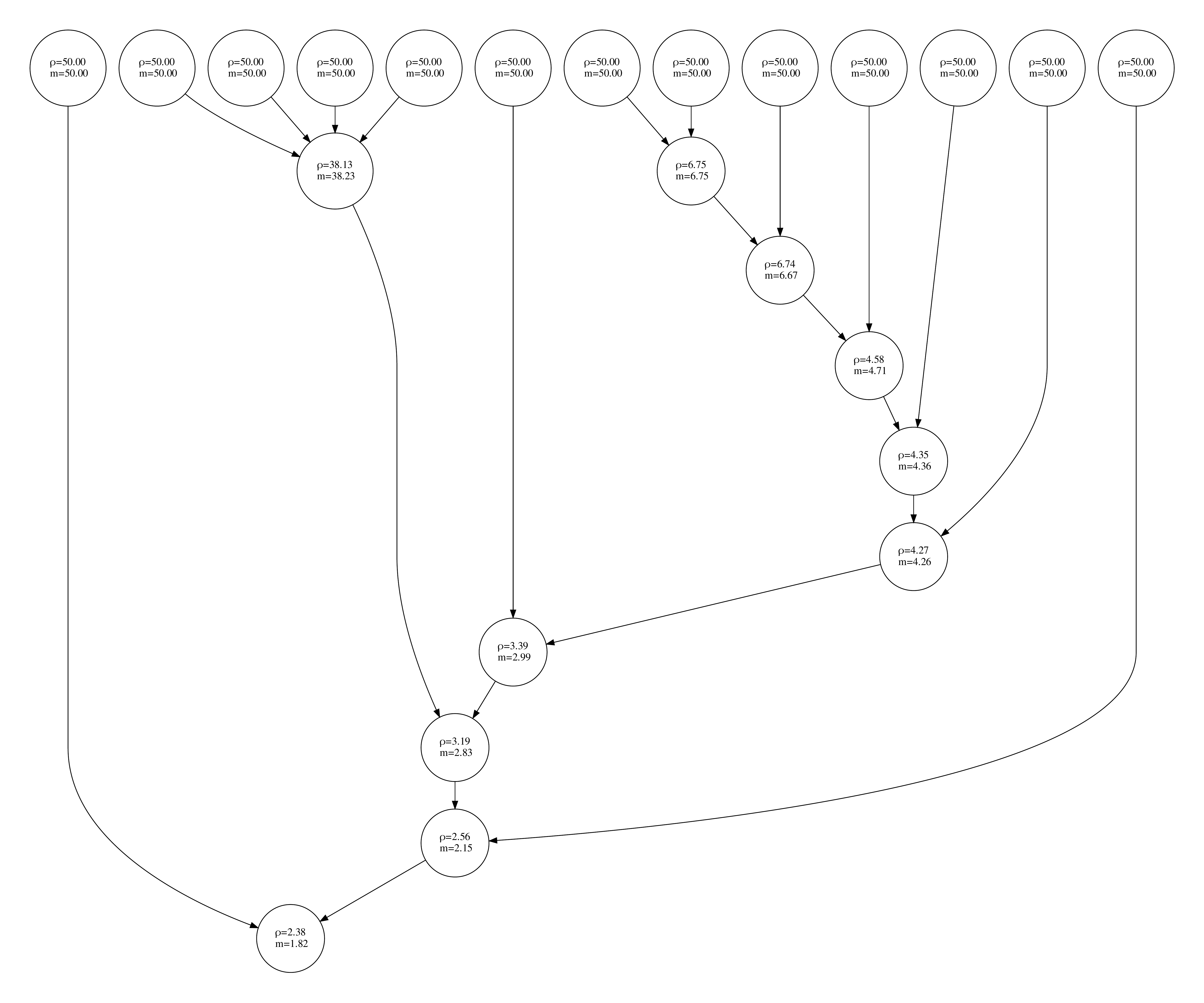}
\centering
\caption{The merging tree for a system with $N=8$ particles is shown. The top row of the tree contains all SHS clusters and the tree follows unique Morse clusters through the continuation process. The trees are the same for each choice of $\kappa$. The values $\rho$ and $m$ in each node give the value of the range parameter for which the merge occurred.
}
\label{N8Merge}
\end{figure*}

\begin{figure*}[!ht] 
\subfigure[]{\includegraphics[width=0.44\textwidth]{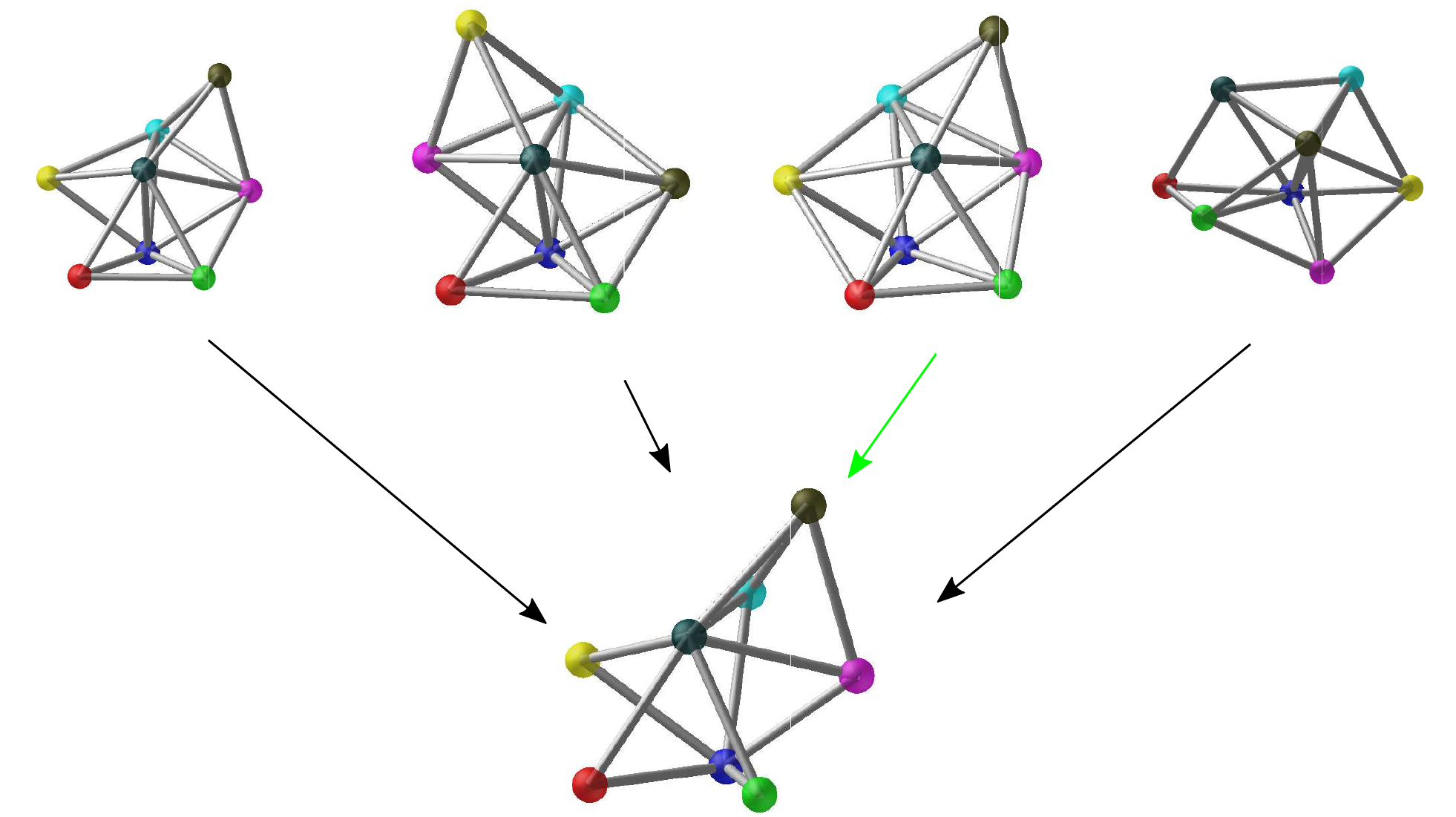}}
\subfigure[]{\includegraphics[width=0.44\textwidth]{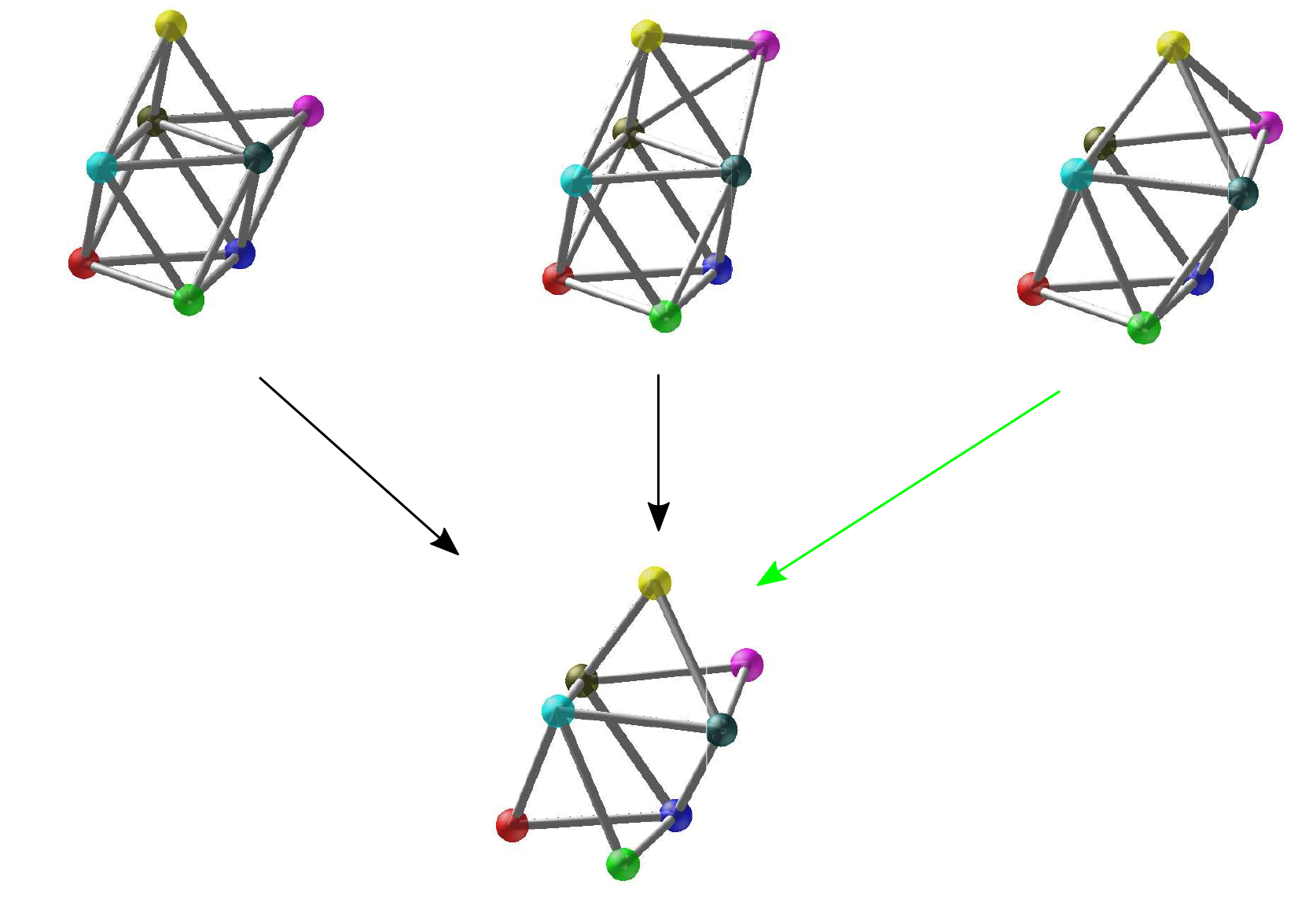}}
\subfigure[]{\includegraphics[width=0.44\textwidth]{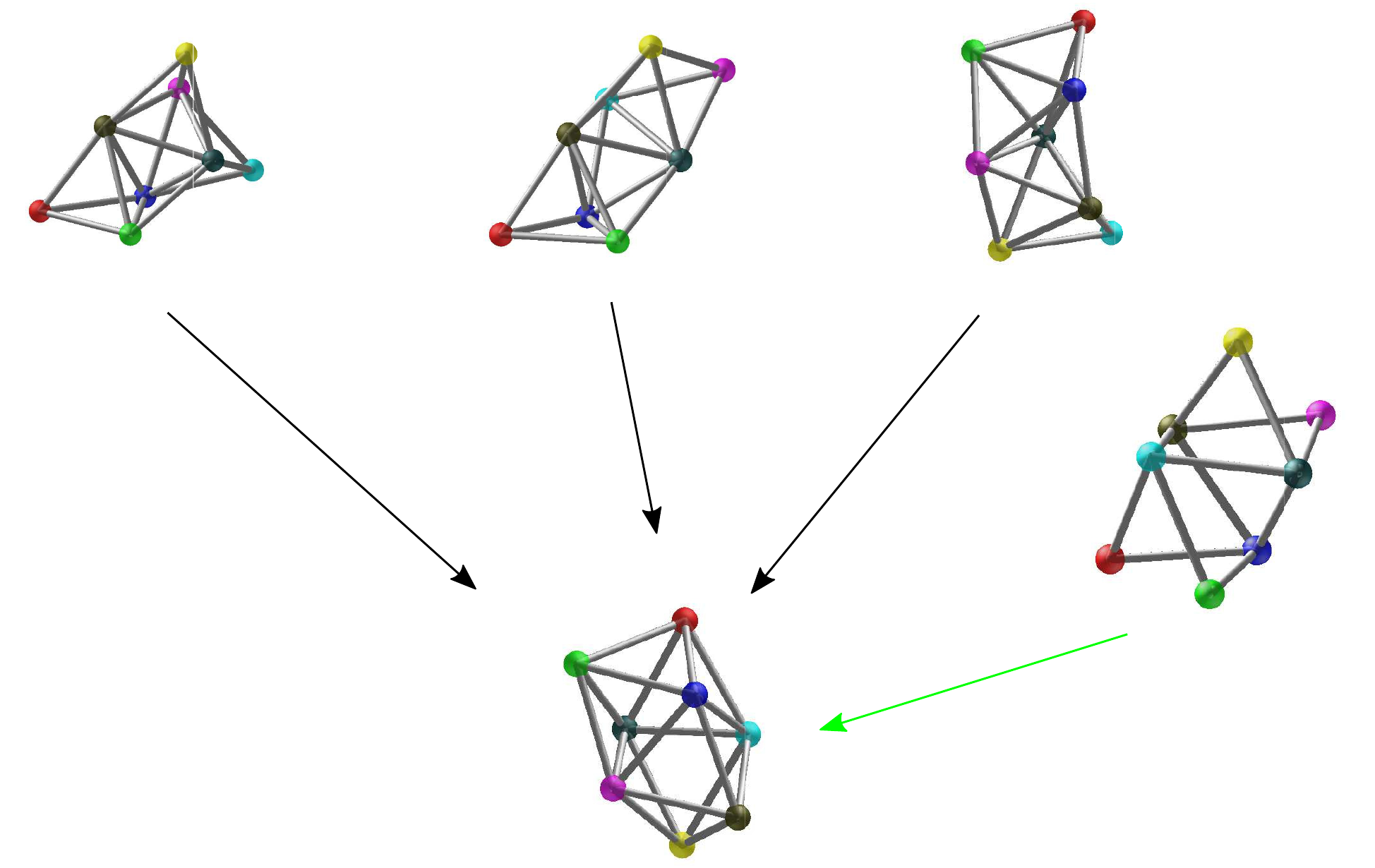}}
\subfigure[]{\includegraphics[width=0.44\textwidth]{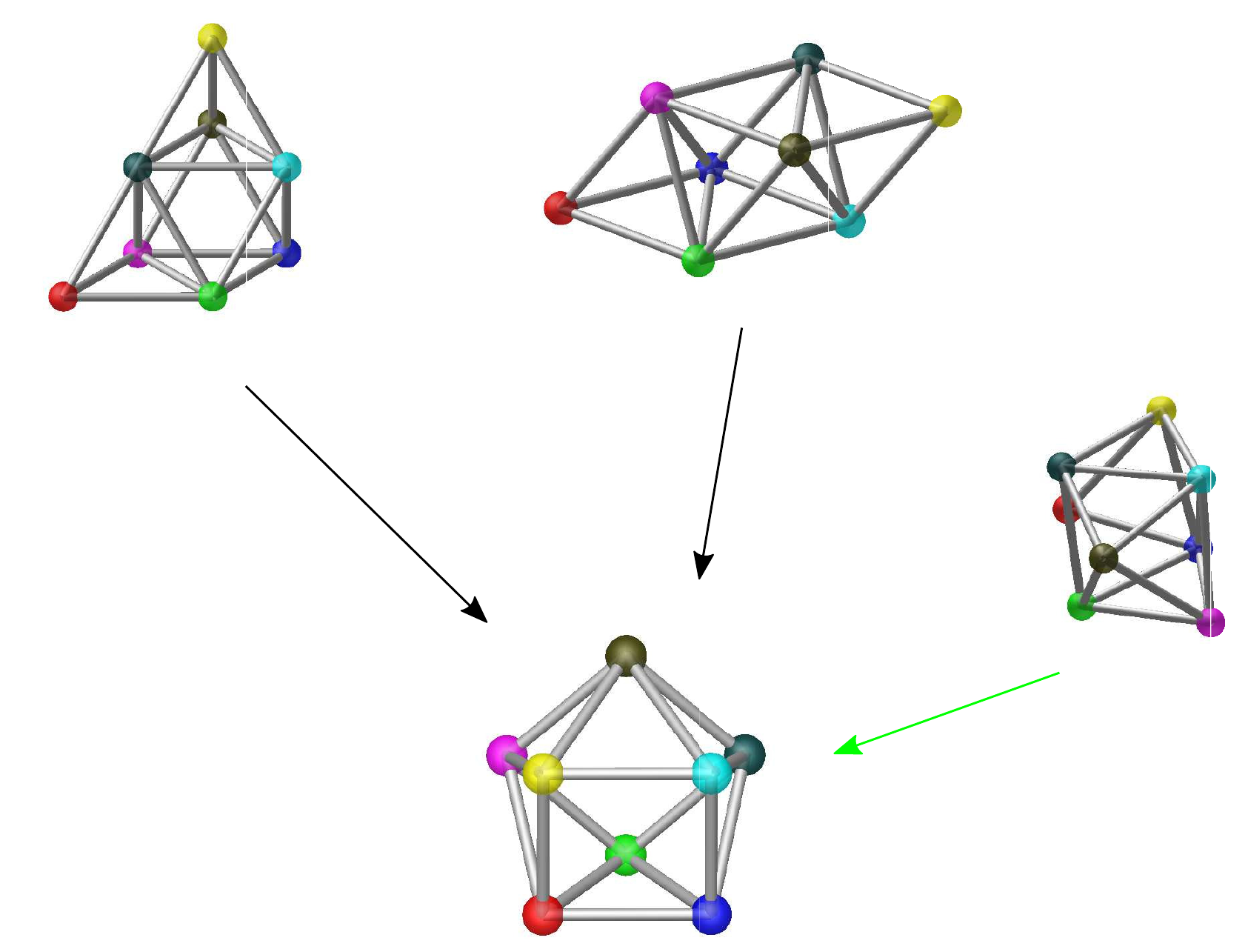}}
\centering
\caption{Cluster images of the major merge events for the $N=8$ system, for values of $\rho$ near (a) $\rho=38$, (b) $\rho=6$, (c) $\rho=4$, and (d) $\rho=2$. The parent clusters are SHS clusters. Green arrows indicate smooth transitions. }
\label{n8clusters}
\end{figure*}

Figure \ref{N8Merge} shows the merging tree for $N=8$. As for $N=7$,  the tree is insensitive to the value of the sticky parameter. Changing the form of the potential affects the range at which clusters merge. Figure \ref{n8clusters} separates the tree into the four major merge events, with SHS clusters at the top and green arrows indicating a smooth transition. In (a), we see the $N=7$, $\rho=38$ mechanism is responsible for this merge. In (b), the re-arranging clusters have an octahedron as a sub-structure, and the smooth cluster is new, i.e. it does not contain any $N=6$ or $N=7$ clusters as sub-structures. In (c), all the SHS clusters have a polytetrahedron as a sub-structure, and they merge with the resulting cluster from (b). Finally (d) contains one structure with an octahedron sub-structure and a new cluster which re-arrange to form the final cluster.

\end{document}